\documentclass[journal]{IEEEtran}
\usepackage[dvips]{graphicx}
\usepackage{epsfig}
\usepackage{amsmath}
\usepackage{epsfig}
\usepackage{array}
\usepackage{amssymb}
\usepackage{color}
\usepackage{verbatim}
\usepackage{float}
\usepackage{multirow}
\newtheorem{remark}{Remark}
\newtheorem{lemma}{Lemma}
\newtheorem{prop}{\textbf{Proposition}}

\begin{document}

\title{Secrecy of Multi-Antenna Transmission with Full-Duplex User in the Presence of Randomly Located Eavesdroppers}

\author
{
	\IEEEauthorblockN
	{
		Ishmam Zabir, \emph{Student Member, IEEE,}
		Ahmed Maksud, \emph{Student Member, IEEE,}
		Gaojie Chen, \emph{Senior Member, IEEE,}
		Brian M. Sadler, \emph{Fellow, IEEE,}
		Yingbo Hua, \emph{Fellow, IEEE}
		\thanks{
		This work was supported in part by the Army Research Office under Grant Number W911NF-17-1-0581. The views and conclusions contained in this document are those of the authors and should not be interpreted as representing the official policies, either expressed or implied, of the Army Research Office or the U.S. Government. The U.S. Government is authorized to reproduce and distribute reprints for Government purposes notwithstanding any copyright notation herein.
		I. Zabir, A. Maksud and Y. Hua (\emph{corresponding author}) are with Department of Electrical and Computer Engineering, University of California, Riverside, CA 92521, USA (email: izabi001@ucr.edu; amaks002@ucr.edu; yhua@ee.ucr.edu.). G. Chen is with School of Engineering, University of Leicester, Leicester, U.K., LE1 7RH (email:gaojie.chen@leicester.ac.uk). B. Sadler is with Army Research Laboratory, Adelphi, MD, USA (email: brian.sadler@ieee.org). }
	}
}

\maketitle

\begin{abstract}
	This paper considers the secrecy performance of several schemes for multi-antenna transmission to single-antenna users with full-duplex (FD) capability against randomly distributed single-antenna eavesdroppers (EDs). These schemes and related scenarios include transmit antenna selection (TAS), transmit antenna beamforming (TAB), artificial noise (AN) from the transmitter, user selection based their distances to the transmitter, and colluding and non-colluding EDs. The locations of randomly distributed EDs and users are assumed to be distributed as Poisson Point Process (PPP). We derive closed form expressions for the secrecy outage probabilities (SOP) of all these schemes and scenarios. The derived expressions are useful to reveal the impacts of various environmental parameters and user's choices on the SOP, and hence useful for network design purposes. Examples of such numerical results are discussed.
\end{abstract}


\section{Introduction}

\label{intro}
Since Wyner's work \cite{wyner}, physical layer security has been studied as an alternative or complementary approach to cryptography for information security. This trend of study has accelerated in recent years given its importance for 5G and future wireless networks \cite{yang}.

Due to the broadcast nature of wireless communications, transmitted information in air is highly vulnerable to eavesdropping unless a positive secrecy rate at the physical layer is achieved.
Many prior works for achieving a positive secrecy rate require that the locations and/or channel-state-information (CSI) of eavesdroppers (EDs or Eve) are known to the legitimate users (also referred to as users) \cite{tekin}-\cite{Yang2017}. This requirement is generally difficult to meet in practice.

One way to handle EDs whose locations and CSI are unknown to users is to assume a statistical model for EDs' CSI where both the small-scale-fading and large-scale-fading of EDs' CSI are statistically modelled. While the small-scale-fading is commonly modelled as Gaussian distributed, the large-scale-fading can be treated by assuming EDs to be randomly distributed according to a Poisson Point Process (PPP) \cite{haenggi2008}-\cite{zheng1}. This paper will also adopt the PPP model to investigate the impact of random EDs locations on secrecy performance which is useful over a time window within which the EDs' locations change randomly. 

%
%

The conventional radio is half-duplex (HD). But full-duplex (FD) radio  promises to be available in the near future \cite{riihonen}-\cite{hua2019}. A user equipped with FD capability can receive a desired information while transmitting an artificial noise (AN) to jam nearby EDs \cite{Chen2017}, \cite{Zheng2017}-\cite{Sohrabi_2019}. We will also refer to this AN as Rx-AN which differs from the AN (along with information signal) transmitted by a multi-antenna transmitter. The latter will also be referred to as Tx-AN. Subject to randomly distributed EDs, schemes based on Tx-AN without Rx-AN have been studied in \cite{Zhang1}-\cite{Zhang2013} for non-colluding EDs and in \cite{Ghogho2011}-\cite{zheng1} for colluding EDs. In \cite{Zhang1}, authors investigated the design of multi-antenna Tx-AN to minimize the secrecy outage probability (SOP) by ignoring thermal noise at EDs. In \cite{zheng2015} and \cite{L Zhang}, authors derived exact closed-form expressions for optimal Tx-AN allocation to minimize SOP. In \cite{zheng3}, authors further investigated secrecy performance under imperfect CSI. The aforementioned studies reveal that Tx-AN (for HD receiver) improves secrecy performance against any EDs' scenarios.

\begin{figure}
	\vspace{-0.1in}
	\begin{center}
		\includegraphics[width = .4\textwidth]{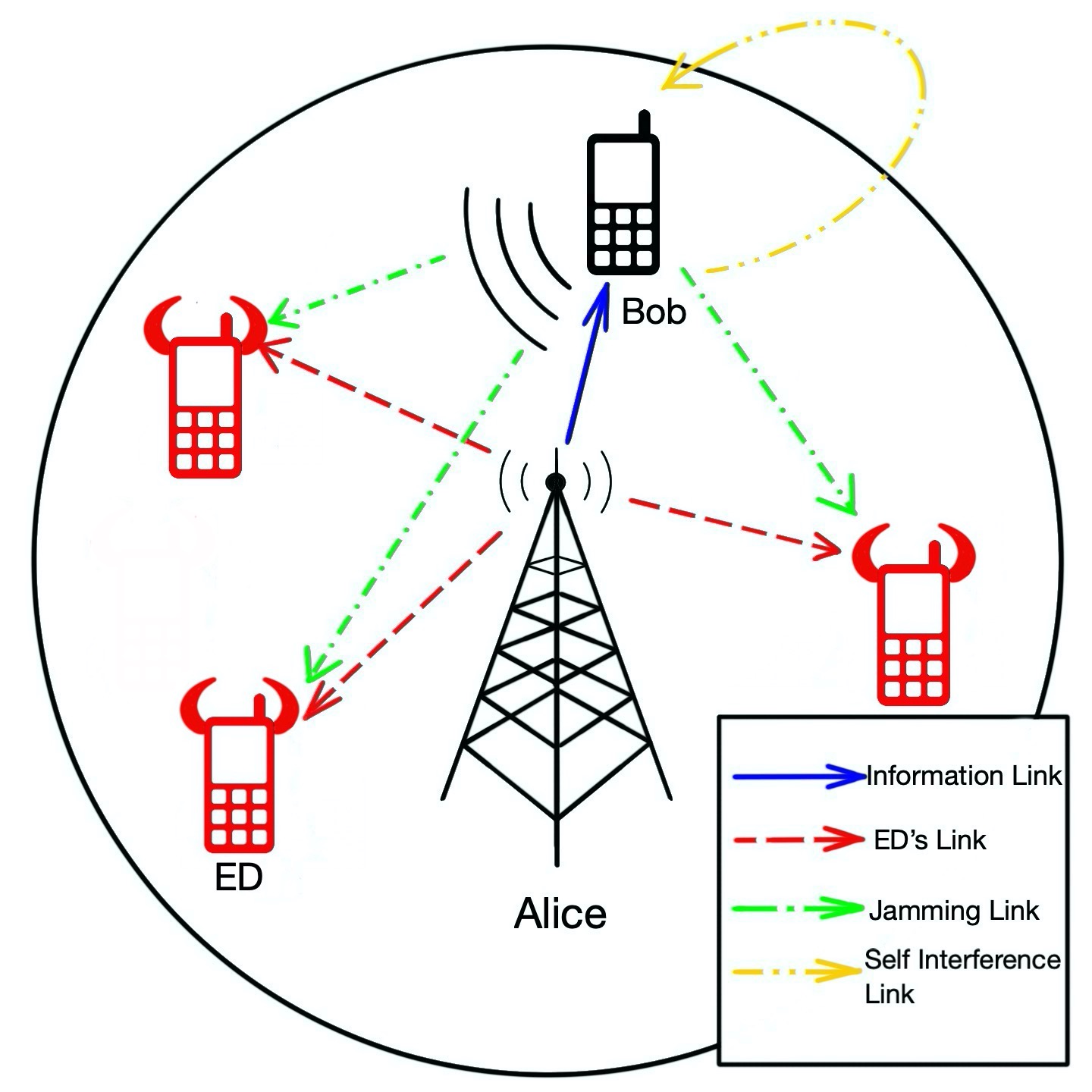}		
		\caption{Wireless network subject to randomly located eavesdroppers where Alice is BS and UE is Bob.}	\label{fig:system_model}
	\end{center}
\end{figure}

This paper will present statistical analyses of SOP for a range of downlink transmission schemes for pairs of multi-antenna base-station (BS) and single-antenna user equipment (UE) in the presence of randomly located EDs, which is illustrated in Fig. \ref{fig:system_model}. These schemes include the following scenarios:
the BS may or may not apply Tx-AN, the UE may or may not apply Rx-AN or equivalently operate in either FD or HD mode, and the EDs may or may not collude with each other to form a virtual antenna array. For randomly distributed UEs, the BS can have them ordered according to their distances to the BS before a downlink transmission may be applied. Furthermore, the BS may apply a transmit-antenna-selection (TAS) scheme or a transmit-antenna-beamforming (TAB) scheme. The TAB scheme requires full CSI knowledge at BS whereas the TAS scheme is a comparatively low-cost low-complexity method \cite{Xie}. In particular, we will focus on the SOP for all the schemes listed above (with exception shown in Table \ref{map_table}). The organization of these analyses is shown in Table \ref{map_table}. Note that HD is a special case of FD, and using no Tx-AN is a special case of using Tx-AN. Much of the mathematical details is given in appendices. Section \ref{simulation} shows numerical results to verify the analysis.  Section IX summarizes the paper.

\begin{table}[ht]
\centering
\caption{Organization of Sections \ref{SOP NC} and \ref{SOP C}}\label{map_table}
\small\addtolength{\tabcolsep}{-3pt}
\begin{tabular}[t]{|c|c|c|c|c|c|}
\hline

\multirow{ 2}{*}{}&\multicolumn{3}{c|}{non-colluding EDs} &\multicolumn{2}{c|}{colluding EDs}\\
\cline{2-6}
&FD UE & HD UE & multi-UE ordering & FD UE & HD UE \\
 \hline
  TAS  &\ref{TAS NC} & \ref{TAS NC HD} & \cite{G.chen_ordering} & \ref{TAS C FD} & \ref{TAS C HD} \\
\hline
  TAB & \ref{TAB NC} & \ref{TAB NC HD} & \ref{TAB ONC} & \ref{TAB C FD}\& \ref{TAB C E} & \ref{TAB C HD}\\

\hline
\end{tabular}
\end{table}%

The key contributions of this paper (a substantial extension of \cite{Ishmam2019}) include the following:
 \begin{itemize}
	\item We derive the closed form expressions of SOP for all the schemes/scenarios listed in Table \ref{map_table}. In the context of randomly distributed EDs, the scheme with both Tx-AN and Rx-AN was not studied before, and none of the schemes listed under colluding EDs was before considered either.
	
	\item  We focus on SOP conditional on user's CSI, which results in a tight lower bound of SOP for both TAS and TAB schemes against randomly located colluding EDs. This is in contrast to \cite{gaojie2017} where TAS was analyzed based on unconditional SOP and zero thermal noise at EDs. The latter is only valid for scenarios of high jamming noise.
	
	\item We extend the analysis shown in \cite{G.chen_ordering} from TAS to TAB with Tx-AN for multiple HD users. Comparisons between TAS and TAB are shown analytically and numerically. (The low cost advantage of antenna selection has been exploited for network throughput as well as physical layer security \cite{Sanayei2004}-\cite{Alves2012}. But TAS shown in \cite{gaojie2017} and \cite{G.chen_ordering} is the most relevant to this paper.)

	\item We reveal the existence of a finite optimum Rx-AN power for both TAS and TAB schemes, which can be also computed based on our closed form SOP expressions.
	

\end{itemize}

The symbols used in this paper are shown in Table \ref{symbol_tabel}.

\begin{table}[ht]
\centering
\caption{Notation and Symbols }\label{symbol_tabel}
\begin{tabular}[t]{|c|c|}
\hline
 Symbol &Definition \\
\hline
 $\mathcal{CN}$ & complex Gaussian\\
 $\Phi$ & the set of locations of EDs\\
 $\rho_{E}$ &  intensity or density of $\Phi$ \\
 $\alpha$ & path loss exponent \\
 $\rho$ & normalized self-interference coefficient \\
 $P_T$ & transmission (signal plus AN) power from Alice\\
 $P_J$ & transmission (jamming) power from Bob\\
 $\epsilon$ & fraction of transmission power at Alice for AN\\
 $\mathbb{E}_{\mathbf{v}}$ & expectation over v\\
 $[x]^{+}$& max(0,x)\\
 $B(x,y)$ & Beta function, where $B(x,y)=\frac{\Gamma(x)\Gamma(y)}{\Gamma(x+y)}$ \\
 $\Gamma(x,y)$ & upper incomplete Gamma function \\
 $\gamma(x,y)$ & lower incomplete Gamma function \\
 $U(a,b,z)$ & Confluent hypergeometric function of
  second kind\\
 $\mathbf{F}(a,b,c;z)$ & Gaussian hypergeometric function \\
 $\mathcal{L}$ & Laplace transform \\
 $\mathbf{E_1}(x)$ & Exponential integral function\\
\hline
\end{tabular}
\end{table}%

\section{SYSTEM MODEL}\label{system model}

We consider a base station (BS or Alice) with multiple antennas located at the center of a circle of radius $R$, which transmits secret information to a single-antenna (omnidirectional) user equipment (UE or Bob). Without loss of generality, we first assume that Bob is located at a unit distance away from Alice. There are randomly located single-antenna (omnidirectional) eavesdroppers (EDs) in the circle, and the random locations of EDs (denoted by $\Phi$) are modeled as a PPP with the intensity $\rho_E$.


The channel gain vector from Alice to Bob is denoted by $\mathbf{h}\in\mathcal{C}^{M\times1}$, which has been normalized to be a complex Gaussian random vector with zero mean and the identity covariance matrix, i.e., $\mathcal{CN}(\mathbf{0},\mathbf{I})$. We assume Bob is equipped with full-duplex antenna (full-duplex can be implemented with either one Tx and one Rx antenna or even with single antenna via RF circulator \cite{Hua2012}) where Bob can transmit and receive at the same time in the same frequency band. The normalized residual instantaneous self-interference channel gain at Bob is $\sqrt{\rho}g_B$ with the distribution $\mathcal{CN}(0,\rho)$ where $\rho$ corresponds to a normalized gain factor (which is relative to the main/user channel gain and should be kept small in application although it can be larger than one if the actual distance between Alice and Bob is relatively large \cite{Hua2017}). The channel vector from Alice to the $e$th ED is $\sqrt{a_e}\mathbf{h}_{AE_{e}} \in \mathcal{C}^{M\times1}$ and distributed as $\mathcal{CN}(\mathbf{0},a_e\mathbf{I})$, and the channel gain from Bob to the $e$th ED is $\sqrt{b_e}{h}_{BE_{e}}$ and distributed as $\mathcal{CN}(0,b_e)$. We also let $a_e=\frac{1}{d_{AE_e}^\alpha}$ with $d_{AE_e}$ being the normalized distance between Alice and the $e$th ED, and $b_e=\frac{1}{d_{BE_e}^\alpha}$ with $d_{BE_e}$ being the normalized distance between Bob and the $e$th ED. Note that $a_e$ and $b_e$ are the large-scale fading parameters as they are dependent on the location of ED while $\mathbf{h}$, $g_B$, $\mathbf{h}_{AE_e}$ and $h_{BE_e}$ are the small-scale fading parameters. We assume that the channels are all quasi-static where the channel coefficients stay constant during transmission of any given packet.

 The secrecy rate of the downlink transmission from Alice to Bob is
\begin{equation}\label{eq:SAB}\small
S_{AB}=[\log_2(1+SNR_{AB})-\log_2(1+ SNR_{AE_*})]^+,
\end{equation}
where $SNR_{AE_*}= \mathcal{F}(SNR_{AEe})$. The operator $\mathcal{F}(.)$ takes the location dependent Signal-to-Noise Ratios (SNRs) of EDs as argument.
The form of $\mathcal{F}(.)$ is dependent on  whether EDs are acting independently or colluding with each other. In the case of non-colluding EDs, the strongest ED channel is considered and the form of $\mathcal{F}(.)$  is defined as
\begin{equation} \small
\mathcal{F}(.)=\max_{e\in \Phi} (.).
\end{equation}
In the case of colluding EDs, we assume that all EDs  can combine their own SNRs to jointly  decode the information bearing signal. We consider passive (distributed) EDs. Since they do not have access to the full CSI between Alice and themselves, they are unable to form a virtual antenna array for colluding. This assumption is  the same as in \cite{haenggi2008}-\cite{zheng1} and \cite{gaojie2017}-\cite{G.chen_ordering}. Thus,
\begin{equation} \small
\mathcal{F}(.)=\sum_{e\in \Phi} (.).
\end{equation}

For a target secrecy rate $R_S$, the SOP is defined as
\begin{equation} \small
P_{out} \overset{\Delta}{=} P(S_{AB}\leq R_S) = P\left [\frac{1+SNR_{AB}}{1+ SNR_{AE_*}}\leq 2^{R_S}\right],
\end{equation}
where $P(\cdot)$ denotes the probability. We will also use $P_{con} \overset{\Delta}{=} 1-P_{out}$.

\subsection{Transmit Antenna Selection}
In the TAS scheme, Alice only transmits via the antenna corresponding to the element in $\mathbf{h}$ that has the largest amplitude. Let $\sqrt{P_T}x_A(k)$ of power $P_T$ be the information signal transmitted from Alice, and $h_{i^*}$ be the element selected from $\mathbf{h}=[h_1,\cdots,h_M]^T$, i.e., $|h_{i^*}|=\max\limits_{i}|h_i|$. Thus, Bob and Eve receive the following signals respectively:
\begin{eqnarray} \small
&y_B(k)=h_{i^*}\sqrt{P_T}x_A(k)+\sqrt{\rho P_J}g_B \tilde w_B(k)+n_B(k), \label{eq AB} \\
&y_{E_e}(k)=\sqrt{a_eP_T}{h_{A_{i^*}E_e}x_A(k)}+\sqrt{b_eP_J}h_{BE_e}w_B(k)\nonumber\\
&+n_{E_e}(k),
\end{eqnarray}
where $\sqrt{P_J}w_B(k)$ of power $P_J$ is the jamming noise or Rx-AN from Bob, $n_B(k)$ and $n_{E}(k)$ are the background Gaussian noises at Bob and Eve each with the unit variance, and the second term of \eqref{eq AB} denotes the residual self-interference. Then, the SNR at Bob is
\begin{equation}\label{eq:SNR_AB_TAS} \small
SNR_{AB}^{TAS}=\frac{|h_{i^*}|^2P_T}{1+\rho |g_B|^2P_J},
\end{equation}
and the SNR at the $e$th Eve is
\begin{equation}\label{eq:SNR_AE_TAS} \small
SNR_{AEe}^{TAS}=\frac{ a_e|h_{A_{i^*}E_e}|^2P_T}{1+b_e|h_{BE_e}|^2P_J}.
\end{equation}

\subsection{Transmit Antenna Beamforming}
 In the TAB scheme, Alice takes the advantage of the complete knowledge of $\mathbf{h}$ by transmitting the following signal:
\begin{equation} \small
 \mathbf{s}(k)=\sqrt{(1-\epsilon)P_T}\mathbf{t}x_{A}(k)+\sqrt{\frac{{\epsilon P_T}}{M-1}}\mathbf{W}\mathbf{v}(k),
\end{equation}
where $x_A(k)$ is the message signal of zero mean and unit variance, $\mathbf{t}=\frac{\mathbf{h}^*}{\|\mathbf{h}\|}$, $\mathbf{W}\in\mathcal{C}^{M\times(M-1)}$ has the orthonormal columns that span the left null space of $\mathbf{t}$ (hence $\mathbf{t}\mathbf{t}^H+\mathbf{W}\mathbf{W}^H=\mathbf{I}$), $\mathbf{v}\in\mathcal{C}^{(M-1)\times 1}$ is the Tx-AN $\mathcal{CN}(\mathbf{0},\mathbf{I})$, and $\epsilon \in\{0,1\}$ is the power fraction factor that splits the total power $P_T$ between the Tx-AN term  and the message term.

Consequently, the received signal at Bob and the $e$th Eve are:
\begin{eqnarray} \small
\begin{aligned}
&y_B(k)=\sqrt{(1-\epsilon)P_T}\|\mathbf{h}\|x_A(k)+\sqrt{\rho P_J}g_B\tilde w_B(k)+n_B(k),\nonumber\\
&y_{E_e}(k)=\sqrt{{a_e}(1-\epsilon)P_T}\frac{\mathbf{h}_{AEe}^T\mathbf{h}^*}{\|\mathbf{h}\|}x_A(k)+\sqrt{b_e P_J}{h}_{BE_e} w_B(k)\nonumber\\
&\;\;\;\;\;\;\;\;\;\;+\sqrt{a_e}\sqrt{\frac{\epsilon P_T}{M-1}}\mathbf{h}_{AEe}^T\mathbf{W}\mathbf{v}(k)+n_{E_e}(k),
\end{aligned}
\end{eqnarray}
respectively. Then the SNR at Bob is
\begin{equation}\label{SNR_bob} \small
SNR_{AB}^{TAB}=\frac{(1-\epsilon){\|\mathbf{h}\|^2} P_T}{1+\rho |g_B|^2 P_J},
\end{equation}
and the SNR at the $e$th Eve is
\begin{eqnarray} \label{eq:SNR_AE_TAB} \small
&&\!\!\!\!\!\!\!\!\!\!\!\!\!\!SNR_{AE_e}^{TAB}=\frac{a_e(1-\epsilon)\frac{|\mathbf{h}_{AE_e}^T\mathbf{h}^*|^2}{\|\mathbf{h}\|^2}P_T}{1+b_e|{h}_{BE_e}|^2P_J+a_e\frac{\epsilon P_T}{M-1}\mathbb{E}_{\mathbf{v}}\{|\mathbf{h}_{AE_e}^T\mathbf{W}\mathbf{v}|^2\}}\nonumber\\
&&\!\!\!\!\!\!\!\!\!\!\!\!\!\!=\frac{a_e(1-\epsilon)\frac{|\mathbf{h}_{AE_e}^T\mathbf{h}^*|^2}{\|\mathbf{h}\|^2}P_T}{1+b_e|{h}_{BE_e}|^2P_J+a_e\frac{\epsilon P_T}{M-1}\|\mathbf{h}_{AE_e}\|^2(1-\frac{|\mathbf{h}_{AE_e}^T\mathbf{h}^*|^2}{\|\mathbf{h}_{AE_e}\|^2\|\mathbf{h}\|^2})}\nonumber\\
&&\!\!\!\!\!\!\!\!\!\!\!\!\!\!=\frac{(1-\epsilon)X_1\Theta P_T}{d_{AE_e}^\alpha+\frac{P_Jd_{AE_e}^\alpha}{d_{BE_e}^\alpha}X_2+\frac{\epsilon P_T}{M-1}X_1(1-\Theta)},
\end{eqnarray}
where $\mathbb{E}_{\mathbf{v}}$ denotes the expectation over $\mathbf{v}$ and $\mathbb{E}_{\mathbf{v}}\{|\mathbf{h}_{AE_e}^T\mathbf{W}\mathbf{v}|^2\}=\mathbb{E}_{\mathbf{v}}\{|\mathbf{h}_{AE_e}^T\mathbf{W}\mathbf{v}\mathbf{v}^T\mathbf{W}^T\mathbf{h}_{AE_e}|^2\}=\mathbf{h}_{AE_e}^T(\mathbf{I}-\mathbf{t}\mathbf{t}^H)\mathbf{h}_{AE_e}$, $X_1=\|\mathbf{h}_{AE_e}\|^2$, $X_2=|h_{BE_e}|^2$ and $\Theta=\frac{|\mathbf{h}_{AE_e}^T\mathbf{h^*}|^2}{\|\mathbf{h}_{AE_e}\|^2\|\mathbf{h}\|^2}$. Note that $X_1$, $X_2$ and $\Theta$ are independent of each other.

Furthermore, $X_1$ has a Chi-squared distribution with $2M$ degrees of freedom (DoF), i.e., its probability density function (PDF) is $f_{X_1}(x)=\frac{x^{M-1}e^{-x}}{\Gamma(M)}$; $X_2$ has a Chi-squared distribution with 2 DoF (also known as the exponential distribution of the unit mean); and $\Theta$ is known to have the beta distribution \cite{hua2019} with parameters $B(1,M-1)$, i.e., $f_{\Theta}(x)=(M-1)(1-x)^{M-2}$. Note that $Beta(a,b)$ distributed random variable $X$ the PDF $f_X(x)=\frac{x^{a-1} (1-x)^{b-1}}{B(a,b)}$.

In order to maintain a data rate $R_D$ from Alice to Bob, we must have $\log_2(1+SNR_{AB}^{TAB})>R_D$, i.e., $1-\epsilon>\frac{{1+\rho|g_B|^2P_J}}{\|\mathbf{h}\|^2P_T}(2^{R_D}-1)$ for the non-negative $\epsilon$.

\subsection{TAB with User Selection (TAB-US)}

In the TAB-US scheme, we assume that Alice (BS) serves multiple single-antenna HD Bobs (UEs) (where $P_J=0$) based on the user's distance from Alice. The locations of Eves and Bobs are all modeled as spatial PPP, i.e., $\Phi_E$ with intensity $\rho_E$ and $\Phi_{U}$ with intensity $\rho_U$ respectively.

Let $d_{AB_n}$ be the distance from Alice to the $n$th (nearest) Bob. Similar to (\ref{SNR_bob}), the SNR at the $n$th Bob is
\begin{equation}\label{} \small
SNR_{AB_n}=\frac{(1-\epsilon)P_T\|\mathbf{h}_{AB_n}\|^2}{ d_{AB_n}^\alpha },
\end{equation}
and, similar to (\ref{eq:SNR_AE_TAB}), the SNR at the $e$th Eve is
\begin{equation}\label{UE_order} \small
SNR_{AE_e}=\frac{a_e(1-\epsilon)\|\mathbf{h}_{AE_e}\|^2\frac{\|\mathbf{h}_{AE_e}^H\mathbf{h}_{AB_n}\|^2}{\|\mathbf{h}_{AE_e}\|^2\|\mathbf{h}_{AB_n}\|^2}P_T}{1+a_e\frac{\epsilon P_T}{M-1}\|\mathbf{h}_{AE_e}||^2(1-\frac{\|\mathbf{h}_{AE_e}^H\mathbf{h}_{AB_n}\|^2}{\|\mathbf{h}_{AE_e}\|^2\|\mathbf{h}_{AB_n}\|^2})},
\end{equation}
where $X_{2,n}=\|\mathbf{h}_{AB_n}\|^2$ is independent from $X_1$ and both follow the Chi-squared distribution with $2M$ degrees of freedom, i.e.,  $f_{X_{2,n}}(x)=f_{X_1}(x)=\frac{x^{M-1}e^{-x}}{\Gamma(M)}$. Also $\Theta=\frac{\|\mathbf{h}_{AE_e}^H\mathbf{h}_{AB_n}\|^2}{\|\mathbf{h}_{AE_e}\|^2\|\mathbf{h}_{AB_n}\|^2}$ follows the $B(1,M-1)$ distribution \cite{hua2019}, and $X_4=\Theta X_1$ is exponentially distributed with mean equal to one. Also, $X_{4,4}=(1-\Theta)X_1$ follows $\Gamma(M-1,1)$ distribution and most importantly $X_{2,n}$, $X_4$ and $X_{4,4}$ are independent.

\section{Secrecy Performance Against Non-colluding Eavesdroppers}\label{SOP NC}
Throughout this section, we study the secrecy performance of both the TAB and TAS schemes against independently acting EDs.
 Furthermore, we analyze the secrecy performance of the TAB scheme as a function of the ordering index of each Bob (among randomly distributed Bobs)  with respect to his distance to Alice.
\subsection{Secrecy performance of the TAS Scheme}\label{TAS NC}
The performance of the TAS scheme was analyzed in \cite{gaojie2017} by assuming that the noise at each node is dominated by the interference. A novelty of the following analysis is an insight that there is generally a nonzero optimal $P_J$. Such an analytical insight would not be possible if the noise is assumed to be negligible from the very beginning of the analysis. Moreover, authors of \cite{gaojie2017} derived the SOP expression averaged over the distribution of legitimate channel. Such analysis does not provide useful insights for a given/common realization of the legitimate channel. In this paper, we study the SOP expression conditioned on the legitimate channel CSI. For a large coherence period of the legitimate channel, the SOP averaged over EDs' distribution can be minimized over the jamming power from FD Bob. Thus, this study enables us to find the optimum allocation at Bob.  We will also show the overall averaged SOP considering the distribution of the legitimate channel.   \\

We will use the following parameterizations: $\beta \overset{\Delta}{=} 2^{R_s}$, $m \overset{\Delta}{=}\frac{P_J}{P_T}$ (``a transmit power ratio''), $Y\overset{\Delta}{=} SNR_{AB}^{TAS}=\frac{|h_{i^*}|^2}{ \frac{1}{P_T}+\rho m|g_B|^2}$ and $Y_0 \overset{\Delta}{=} \frac{Y}{\beta}+\frac{1}{\beta}-1$. Note that for any given realization of $\mathbf{h}$ and $g_B$, $Y$ is a given constant. Hence, $P[S_{AB}^{TAS}>R_s|\Phi,\mathbf{h},g_B]=P[S_{AB}^{TAS}>R_s|\Phi,Y]$. \\

\begin{prop}\label{TASprop}
 Conditioned on $\mathbf{h}$ and $g_B$, the probability of achieving a secrecy rate strictly larger than $R_s$ using the TAS scheme is given by
\begin{eqnarray} \label{eq:Piny} \small
P_{con,Y}&=&\exp\bigg[-\rho_E\int_{0}^{R}\int_{0}^{2\pi}\Psi(Y,r,\theta)r\mbox{d}\theta \mbox{d}r\bigg],
\end{eqnarray}
where
\begin{equation}\label{introduce_psi} \small
\Psi(Y,r_e,\theta_e)=\frac{\exp(-\frac{{d_{AE_e}^{\alpha}}}{P_T}Y_0)}{1+m(\frac{d_{AE_e}}{d_{BE_e}})^{\alpha}Y_0},
\end{equation}
and $(r_e,\theta_e)$ are the polar coordinates of the location of the $e$th Eve with the origin at the location of Alice. Also  $d_{AE_e}=r_e$ and $d_{BE_e}=\sqrt{r_e^2+ d^2-2r_e\cos\theta}$.
\end{prop}
The proof is shown in Appendix \ref{sec:TAS}.
\begin{remark}
It is obvious that $P_{con,Y}$ is a decreasing function of $\Psi(Y,r,\theta)$. One can verify the following statements subject to $P_T>0$:
\begin{itemize}
	\item If $R_s=0$, then $\Psi(Y,r,\theta)$ is invariant to $P_T$.
	\item If $R_s=0$ and the product $\rho P_J$ is a fixed constant, then as $P_J$ increases to $\infty$, $\Psi(Y,r,\theta)$ decreases monotonically to zero and hence $P_{con,Y}$ increases monotonically to one.
	\item If $\rho\ll1$, then in a region of small $P_J$, $Y$ and hence $Y_0$ are approximately invariant to $P_J$. But in this case, $\Psi(Y,r,\theta)$ decreases as $P_J$ increases (since $Y_0\left(\frac{d_{AE}}{d_{BE}}\right)^\alpha$ is not small) and hence $P_{con,Y}$ increases as $P_J$ increases.
\end{itemize}
\end{remark}

\begin{table}[t]
\centering
\caption{Effects of parameters on $\Psi(Y,r,\theta)$ \& $P_{con,Y}$ when $R_s=0$, where $-$, $\uparrow$ and $\downarrow$ denote invariance, increasing and decreasing, respectively. }\label{Psi table}
\small\addtolength{\tabcolsep}{-5pt}
\begin{tabular}[t]{|c|c|c|c|c|}
\hline
   & $P_T$ & $P_J \to \infty(\rho P_J=a)$ & $P_J \uparrow (\leq {P_J }^*)$ &$P_J \uparrow (\geq {P_J }^*)$ \\
\hline
  $\Psi(Y,r,\theta)$  & $-$ & $\to 0$ & $\downarrow$ & $\uparrow$\\
\hline
  $P_{con,Y}$  & $-$ & $\to 1$ & $\uparrow$ & $\downarrow$\\

\hline
\end{tabular}
\end{table}%

The aforementioned statements are summarized in Table \ref{Psi table} where $ {P_J }^*$ is nonzero optimal value of $P_J$ and $a$ is an arbitrary constant. Here $P_J \to \infty$ implies that the jamming power from Bob is large or more precisely $P_J\gg\frac{1}{\rho|g_B|^2}$. The expression \eqref{eq:Piny} provides the relationship between the target secrecy rate $R_s$ and different parameters in the network. To obtain $P_{con,Y}$ numerically, the double integrals shown there need to be computed for a given choice of the path loss exponent $\alpha$. In general, experimentally estimated $\alpha$ results in difficulty for simplification of the double integrals. But for $R_s=0$ (i.e. $Y_0=Y$) and $\alpha=2$, a simplification can be shown to be
\begin{align}
P_{con,Y}&=\exp\bigg[-\rho_E \bigg(\frac{\pi P_T}{Y}\big(1-\exp(-\frac{YR^2}{P_T})\big)-\pi mY\nonumber\\
&\times\int_{0}^{R^2}\frac{\exp(-\frac{Yr}{P_T})r}{\sqrt{\big((1+mY)r+d^2\big)^2-4rd^2}} \mbox{d}r \bigg) \bigg]
\end{align}
which is shown in Appendix \ref{sec:simplification}.

Then, the unconditional $P_{con}=\mathbb{E}_Y[P_{con,Y}]$ can be obtained by
\begin{equation}\label{eq:Pin} \small
P_{con}=P[S_{AB}^{TAS}>R_s]=\int_{y=0}^\infty P_{con,y}f_Y(y)\mbox{d}y,
\end{equation}
where  the distribution of $Y$ due to the random $|h_{i^*}|^2$ and $|g_B|^2$ is given in the following lemma:
\begin{lemma}\label{Lemma1}
	The cumulative distribution function (CDF) of $Y$ is
	\begin{equation}\label{eq:18} \small
	F_Y(y)=\sum_{i=0}^{M}C_{i}^M(-1)^i\frac{e^{-\frac{iy}{P_T}}}{1+iy\rho m},
	\end{equation}
\end{lemma}
where $C^M_i=\frac{M!}{(M-i)!i!}$. Hence the PDF of $Y$ is
\begin{equation}
f_Y(y)=\sum_{i=0}^{M}C_{i}^M(-1)^{i+1}i{e^{-\frac{iy}{P_T}}}\frac{(\frac{iy\rho m}{P_T}+\frac{1}{P_T}+\rho m)}{(1+iy\rho m)^2}.
\end{equation}

Next, we consider $P_{con,Y}$ in the two special cases: $P_J=0$ and $P_J \to \infty$.

\subsubsection{The Case of $P_J=0$} \label{TAS NC HD}
Now we consider the case of $P_J=0$ thus $m=0$ and assume that $P_T\gg\frac{1-\beta}{|h_{i^*}|^2}$ thus $Y_0\approx\frac{P_T|h_{i^*}|^2}{\beta}$. It follows from (\ref{introduce_psi}) that $\Psi(Y;r,\theta)=\exp(-\frac{d_{AEe}^\alpha}{P_T}Y_0)=\exp\left(-r^\alpha\frac{|h_{i^*}|^2}{\beta}\right)$. Hence
\begin{eqnarray}\label{eq:HD_TAS} \small
\frac{\ln P_{con,Y}}{\rho_E}&=&-\int_{0}^{R}\int_{0}^{2\pi}\Psi(Y;r,\theta)r\mbox{d}\theta \mbox{d}r\nonumber\\
&=&-2\pi\int_0^R\exp\left(-r^\alpha\frac{|h_{i^*}|^2}{\beta}\right)r\mbox{d}r\nonumber\\
&=&-\frac{2\pi{\beta}^{\frac{2}{\alpha}}}{\alpha(|{h_{i^*}}|^2)^{\frac{2}{\alpha}}}\int_{0}^{\frac{R^{\alpha}(|{h_{i^*}}|^2}{\beta}}
\exp(-z)z^{\frac{2}{\alpha}-1}\mbox{d}z\nonumber \\
&=&-\frac{2\pi{\beta}^{\frac{2}{\alpha}}}{\alpha(|{h_{i^*}}|^2)^{\frac{2}{\alpha}}}\mathbf{\gamma}\left(\frac{2}{\alpha},\frac{|{h_{i^*}}|^2 R^{\alpha}}{\beta}\right),
\end{eqnarray}
where $z=r^\alpha\frac{|h_{i^*}|^2}{\beta}$ and $\gamma(x,y)=\int_0^y z^{x-1}e^{-z}\mbox{d}z$ is the lower incomplete gamma function which increases monotonically with $y$. From \eqref{eq:HD_TAS}, it is clear that $P_{con,Y}$ monotonically decreases as $R$ increases. In particular,
\begin{equation}\label{eq:limit} \small
\lim_{R\rightarrow\infty}\frac{\ln P_{con,Y}}{\rho_E}=-\pi\left (\frac{\beta}{|{h_{i^*}}|^2}\right )^{\frac{2}{\alpha}}
\frac{2}{\alpha}\Gamma\left(\frac{2}{\alpha}\right),
\end{equation}
where $\Gamma(x)=\int_0^\infty z^{x-1}e^{-z}\mbox{d}z$. It is known that $x\Gamma(x)=\Gamma(x+1)$ for positive $x$ and $\Gamma(x+1)$ decreases to one as $x$ decreases to zero. Then, provided $\frac{\beta}{|h_{i^*}|^2}>1$, the above limit increases as $\alpha$ increases. The result \eqref{eq:limit} serves as a benchmark corresponding to a HD Bob.

\subsubsection{The case of $P_J= \infty$}
We now consider the case of $P_J=\infty$ and also assume $R_s=0$ and $\alpha=2$. In this case, $Y_0=Y=0$ and $mY=\frac{|h_{i^*}|^2}{\rho|g_B|^2}$. Then, following a similar derivation as that in section 1 of the supplement,
one can verify that
\begin{align}\label{eq:PJ_infty_TAS} \small
&\frac{\ln P_{con,Y}}{\rho_E}=-{2\pi}\int_{r=0}^{R}\bigg(1- \frac{1}{{\sqrt{1+\rho\frac{|g_B|^2}{|{h_{i^*}}|^2}(1+\frac{d}{r})^2}}}\nonumber\\
&\times\frac{1}{{\sqrt{1+
\rho\frac{|g_B|^2}{|{h_{i^*}}|^2}(1-\frac{d}{r})^2 }}}\bigg)r\mbox{d}r,
\end{align}
where the integrant converges to $\left(1-\frac{1}{1+\rho \frac{|g_B|^2}{|h_{i^*}|^2}}\right)r$ as $r$ becomes large and the integral goes to $\infty$ as $R\rightarrow\infty$. Hence $\lim_{R\rightarrow\infty}P_{con,Y}=0$. This result suggests that $P_J$ should not be too large. Combining this with a previous result for small $P_J$ implies that there is generally a finite nonzero optimal $P_J$.

\subsection{Secrecy performance of the TAB Scheme}\label{TAB NC}

Unlike the TAS scheme, Alice will now use all transmit antennas via beamforming to transmit each  information symbol. We will assume that all the channel links from Alice to Bob are independent and identically distributed.

In addition to $m=\frac{P_J}{P_T}$ and $\beta=2^{R_s}$, we will use $Z=\frac{SNR_{AB}^{TAB}}{(1-\epsilon)}=\frac{\|\mathbf{h}\|^2}{\frac{1}{P_T}+\rho m |g_B|^2}$, $C=\frac{Z}{\beta P_T}-\frac{(1-\frac{1}{\beta})}{(1-\epsilon)P_T }$, $f_e=(\frac{d_{AE_e}}{d_{BE_e}})^{\alpha}m$ (``a large scale receive power ratio'') and $G=\frac{1}{C P_T} = \frac{\beta(1+\rho P_J|g_B|^2)}{P_T\|\mathbf{h}\|^2}
$. The random variables $Z$, $C$ and $G$ are one-to-one related to each other. We will use $z$, $c$ and $g$ for the realizations of $Z$, $C$ and $G$ respectively. Unlike $f_e$, the variables $z$, $c$ and $g$ are invariant to the locations of Eves but dependent on the small scale fading parameters $\mathbf{h}$ and $g_B$.  For given realization of $\mathbf{h}$ and $g_B$, $z$ is  given. For $m=0$, $Z=P_T\|\mathbf{h}\|^2$ has obviously a Chi-squared distribution with $2M$ DoF. For $m> 0$, one can prove the following lemma:
\begin{lemma}\label{lemma2}
	If $m> 0$, the legitimate channel's $SNR_{AB}$ (which is $Z$) has the following PDF (shown in Appendix \ref{lemma2proof})
	\begin{equation}\label{eq:fZz} \small
	f_Z(z)=\frac{M\rho m(z\rho m)^{M-1}}{(1+z\rho m)^{M+1}}e^{\frac{1}{\rho P_J}}.
	\end{equation}
\end{lemma}

\begin{prop}\label{TABprop}
Conditioned on $\mathbf{h}$ and $g_B$, the probability of achieving a secrecy rate strictly larger than $R_s$ using the TAB scheme is given by
\begin{align}\label{eq:TAB} \small
&P_{con,\mathbf{h},g_B}=P_{con,z}=\exp\bigg[-\rho_E\int_{0}^{R}r\int_{0}^{2\pi}\Omega(\frac{1}{{g}};r,\theta)\mbox{d}\theta \mbox{d}r\bigg],
\end{align}
and hence $P_{con}=\int_0^\infty P_{con,z}f_Z(z)dz$
where
\begin{equation}\label{OMEGA} \small
\Omega(\frac{1}{{g}};r,\theta)=\frac{e^{\frac{d_{BE}^\alpha}{P_J}}}{(1+\frac{f_e}{{g}})(1+\frac{\epsilon}
{(M-1){g}} )^{M-1}}
\end{equation}
and all other variables are defined before.
\end{prop}

The proof is shown in Appendix \ref{sec:TAB}.

\begin{remark}
From \eqref{eq:TAB} and \eqref{OMEGA}, one can also verify the following subject to $P_T>0$:
\begin{itemize}
	\item For  $\epsilon>0$,  $P_{con,\mathbf{h},g_B}\rightarrow 1$ as $P_T\rightarrow\infty$. $(1+\frac{\epsilon}{(M-1){g}})^{M-1}$ converges to $e^{\epsilon/g}$ as $M$ increases. For large $P_T$, $\frac{1-\frac{1}{\beta}}{(1-\epsilon)P_T}\approx0$ so, $\frac{f_e}{g}=\left(\frac{d_{AEe}}{d_{BEe}}\right)^\alpha\frac{\|\mathbf{h}\|^2}{\beta(\frac{1}{P_J}+
\rho|g_B|^2)}$ which is invariant to $P_T$ and $\frac{\epsilon}{g}=\frac{\epsilon P_T\|\mathbf{h}\|^2}{\beta(1+\rho P_J|g_B|^2)}$ which goes to $\infty$ as $P_T\rightarrow\infty$ .
	
\item $P_{con,\mathbf{h},g_B}$ increases as $\rho$ decreases. As $\rho$ decreases, $Z$ and $c$ increase, and hence $g$ and $\Omega(\frac{1}{{g}};r,\theta)$ decrease,  and hence $P_{con,\mathbf{h},g_B}$ increases.

\item If $\rho \to 0$, then $P_{con,\mathbf{h},g_B}\to 1$ as $P_J\to \infty$.
	\item If  $\epsilon=0$, the optimal $P_J$ is $\infty$. If $\epsilon=0$ then it follows  from \eqref{eq:TAB} and \eqref{OMEGA} that
	\begin{align}\label{eq:no AN} P_{con,\mathbf{h},g_B}&= \exp\bigg[-\rho_E\int_{0}^{R}r\int_{0}^{2\pi}\frac{e^{\frac{d_{BE_e}^\alpha}{P_J}}}
{1+\frac{f_e}{{g}}}\mbox{d}\theta \mbox{d}r\bigg],
	\end{align}
which is independent of $P_T$ and monotonically increases as $P_J$ increases. Thus the optimum $P_J$ is $\infty$.
	
	\item For $\epsilon>0$ and $P_T>0$, the optimal $P_J$ is a finite positive number. For $\epsilon>0$ and $P_T>0$, $\frac{f_e}{g}$ monotonically increases to $\frac{\|\mathbf{h}\|^2}{\beta\rho|g_B|^2}$ as $P_J\rightarrow\infty$ and $1+\frac{d_{BE_e}^\alpha}{P_J}$ monotonically decreases to $1$ as $P_J\rightarrow\infty$. So, $\frac{1+\frac{d_{BE_e}^\alpha}{P_J}}{1+\frac{f_e}{{g}}}$ monotonically decreases to $\frac{1}{(1+\frac{\|\mathbf{h}\|^2}{\beta\rho|g_B|^2})}$ for $P_J\rightarrow\infty$. We can also observe that $\frac{\epsilon}{g}$ monotonically decreases to $0$ for $P_J\rightarrow\infty$, so $\frac{1}{(1+\frac{\epsilon}{(M-1)g})^{M-1}}$ monotonically increases to $1$ as $P_J\rightarrow\infty$. So, we can conclude that there is a finite positive $P_J$ at which $P_{con,\mathbf{h},g_B}$ is maximized.\\
	
	\item For $R_s=0$, $\Omega(\frac{1}{{g}};r,\theta)$ is a decreasing function of $\epsilon$ which makes the upper bound of $\epsilon$ optimal. Furthermore, $\Omega(\frac{1}{{g}};r,\theta)$ is rather flat around the optimal $P_J$, which makes it easy to find a practically optimal $P_J$.
\end{itemize}
\end{remark}

The aforementioned observations are summarized in Table \ref{Omega table}.

\begin{table}[t]
\centering
\caption{Effects of parameters on $\Omega(z,r,\theta)$ \& $P_{con,z}$ when $R_s=0$}\label{Omega table}
\small\addtolength{\tabcolsep}{-4pt}
\begin{tabular}[t]{|c|c|c|c|c|}
\hline
   & $P_T \to \infty$ & $P_J \to \infty$ & $P_J \uparrow( \leq {P_J }^*)$ &$P_J \uparrow ( \geq {P_J }^*)$ \\
\hline
  $\Omega(z,r,\theta)|_{ \epsilon=0}$  & $-$ & $\downarrow\to const.$ & $\downarrow$ & $-$\\

  \hline
  $P_{con,z}, \epsilon=0$  & $-$ & $\uparrow\to const.$ & $\uparrow$ & $-$\\
  \hline
  $\Omega(z,r,\theta)|_{ \epsilon \not = 0}$  & $\to 0$ & $\uparrow\to const.$ & $\downarrow$ & $\uparrow$\\
\hline
  $P_{con,z}, \epsilon \not = 0$  & $\to 1$ & $\downarrow\to const.$ & $\uparrow$ & $\downarrow$\\

\hline
\end{tabular}
\end{table}%

As shown in Appendix \ref{proof of convexity}, $\Omega(\frac{1}{{g}};r,\theta)$ for any $r$ and $\theta$ is a unimodal function with its minimum at a finite positive value of $P_J$. Therefore, $\int_{0}^{R}r\int_{0}^{2\pi}\Omega(\frac{1}{{g}};r,\theta)\mbox{d}\theta \mbox{d}r$ must also have its minimum at a finite positive value of $P_J$, or equivalently $P_{con,z}=\exp(-\rho_E\int_{0}^{R}r\int_{0}^{2\pi}\Omega(\frac{1}{{g}};r,\theta)\mbox{d}\theta \mbox{d}r)$ has its peak at that value of $P_J$.


Next, we consider two special cases for which the double integral in \eqref{eq:TAB}
 can be simplified.

\subsubsection{Bob in Full Duplex Mode with $\beta=1$ and $\alpha=2$}

For $\beta=1$ and $\alpha=2$, it is shown in Appendix \ref{sec:TAB_alpha=2} that
\begin{align}\label{eq:TAB:alpha=2} \small
&\int_{0}^{R}\int_{0}^{2\pi}\Omega(z;r,\theta)\mbox{d}\theta r\mbox{d}r=\frac{2\pi}{(1+\frac{z\epsilon}{M-1})^{M-1}}\nonumber \\
&\qquad\times\int_{0}^{R}\bigg(1-\frac{1}{\sqrt{1+\frac{(r+d)^2}{r^2zm}}\sqrt{1+\frac{(r-d)^2}{r^2 zm}}}\bigg)r\mbox{d}r.
\end{align}
And $P_{out}=1-P_{con}$ versus $P_J$ and $\epsilon$ will be illustrated in Fig. \ref{fig:eps_varying} from which we will see that for a given $\epsilon$ there is an optimal $P_J$ and the optimal $P_J$ is not very sensitive to $\epsilon$.

Furthermore,
if $P_J\to\infty$, then $z \to 0$, $zm \to\frac{\|\mathbf{h}\|^2}{\rho|g_B|^2}$ and (\ref{eq:TAB:alpha=2}) yields
\begin{align}\label{eq:PJ_infty_TAB} \small
&\int_{0}^{R}\int_{0}^{2\pi}\Omega(z;r,\theta)d\theta rdr={2\pi}\int_{0}^{R}\bigg(1-\nonumber\\
&\qquad \frac{1}{{\sqrt{1+\rho\frac{|g_B|^2}{\|\mathbf{h}\|^2}(1+\frac{d}{r})^2}}{\sqrt{1+\rho\frac{|g_B|^2}
{\|\mathbf{h}\|^2}(1-\frac{d}{r})^2}}}\bigg)r\mbox{d}r.
\end{align}
Comparing \eqref{eq:PJ_infty_TAS} and \eqref{eq:PJ_infty_TAB}, we see a similar structure of the two expressions. Since $\|\mathbf{h}\|^2\geq\max\limits_{i\in M}|h_i|^2$, the TAB scheme always yields a lower SOP than the TAS scheme.

Also note that if $\epsilon > 0$ and $P_T\to \infty$, then \eqref{eq:TAB:alpha=2} implies that the SOP of the TAB scheme becomes one (similar to the case for TAS).

\subsubsection{Bob in Half-Duplex Mode} \label{TAB NC HD}

In this case, we have $P_J=0$ and $z=P_T \|\mathbf{h}\|^2$. Also assuming a large $P_T$, it is shown in Appendix \ref{sec:HD_TAB} that
\begin{align}\label{eq:HD_TAB1} \small
&\int_{0}^{R}\int_{0}^{2\pi}\Omega(z;r,\theta)\mbox{d}\theta r\mbox{d}r\nonumber\\
&=\frac{2\pi{\beta}^{\frac{2}{\alpha}}}{\alpha(\|\mathbf{h}\|^2)^{\frac{2}{\alpha}}(1+\frac{\epsilon P_T}{M-1}\frac{\|\mathbf{h}\|^2}{\beta})^{M-1}}\mathbf{\gamma}\big(\frac{2}{\alpha}, \frac{R^{\alpha} \|\mathbf{h}\|^2}{\beta} \big).
\end{align}

Here $\mathbf{\gamma}(\frac{2}{\alpha},\frac{R^{\alpha}\|\mathbf{h}\|^2}{\beta})$ is the lower incomplete gamma function and increases monotonically as $R$ increases. \eqref{eq:HD_TAB1} is similar to \eqref{eq:HD_TAS} and is independent of $P_T$ when $\epsilon=0$. Since $\|\mathbf{h}\|^2 \geq\max\limits_{i\in M}|{h_i}|^2$, the HD-TAB (even without using AN) results in a better secrecy performance than the HD-TAS. Note that  the secrecy performance of the HD-TAB depends on $P_T$ when $\epsilon>0$. Furthermore, the term $\int_{0}^{R}\int_{0}^{2\pi}\Omega(z;r,\theta)\mbox{d}\theta r\mbox{d}r$ is inversely proportional to the factor $\left(1+\frac{\epsilon P_T\|\mathbf{h}\|^2}{(M-1)g}\right)^{M-1}$. Thus, the term $\int_{0}^{R}\int_{0}^{2\pi} \Omega(z;r,\theta)\mbox{d}\theta r\mbox{d}r$ and hence SOP decreases as the number of transmit antenna $M$ increases.

\subsection{Secrecy performance of the TAB-US Scheme}\label{TAB ONC}

In \cite{G.chen_ordering}, a TAS based downlink transmission scheme for multiple ordered half-duplex receivers or ``a TAS based User Selection (US) scheme'' was considered. In this section, we consider a TAB based counter part of the above scheme, which will be referred to as the TAB-US  scheme.

As shown in Appendix \ref{sec:TAB_ordering}, we have
\begin{prop}\label{TAB_odering_e}
For $\epsilon \geq 0$, the probability of achieving a secrecy rate strictly larger than $R_s$ conditional on the distance of a selected user is
\begin{eqnarray}\label{eq:ordering} \small
   P[S_{AB}>R_s|d_{AB_n}]&=\exp\bigg[ \frac{-2\pi\rho_E }{\alpha} \frac{({\beta d_{AB_n}^\alpha})^MB(M-\frac{2}{\alpha},\frac{2}{\alpha})}{(\frac{\epsilon P_T}{M-1})^{M-\frac{2}{\alpha}}} \nonumber \\
  \times &U(M-\frac{2}{\alpha},2-\frac{2}{\alpha}, \frac{(M-1)\beta d_{AB_n}^\alpha}{\epsilon P_T})  \bigg]
\end{eqnarray}
where $d_{AB_n}$ is the distance between Alice and the $n$th closest user, and $U$ denotes the confluent hypergeometric function of the second kind \cite{Confluent}.
\end{prop}

With $P(S_{AB}>R_s|d_{AB_n})$ and $f_{d_{AB_n}}(x)$ from lemma \ref{d_AB} in Appendix \ref{sec:TAB_ordering0}, one can readily compute the SOP $P(S_{AB}<R_s)=\int_0^\infty P(S_{AB}>R_s|x)f_{d_{AB_n}}(x)dx$ for any $\epsilon$. We will show via simulation that the TAB-US scheme outperforms the TAS-US scheme.

As shown in Appendix \ref{sec:TAB_ordering0}, for the special case of $\epsilon=0$, $P(S_{AB}<R_s)$ can be simplified into:
\begin{align}\label{ep=0_UE} \small
P[S_{AB}>R_s]=\frac{1}{\bigg(1+\frac{\rho_E}{\rho_U}\frac{2}{\alpha}\beta^{\frac{2}{\alpha}}
B(M-\frac{2}{\alpha},\frac{2}{\alpha}) \bigg)^n},
\end{align}
where $\frac{\rho_E}{\rho_U}$ is the ratio of the density of EDs over that of the legitimate receivers.

Furthermore,
for $n=1$ (the nearest Bob), \eqref{ep=0_UE} reduces to
\begin{eqnarray} \small
P[S_{AB}<R_s]=1-\frac{1}{1+\frac{\rho_E}{\rho_U}\frac{2}{\alpha}\beta^{\frac{2}{\alpha}}
B(M-\frac{2}{\alpha},\frac{2}{\alpha}) }.
\end{eqnarray}

\section{Secrecy Performance Against Colluding Eavesdroppers} \label{SOP C}

In this section, we consider the situation that EDs can share all information  to decode the message. Since Alice knows the channel between Alice and Bob, the secrecy performance conditional on $\mathbf{h}$ and $g_B$ is a useful measure. In one coherence period, $\mathbf{h}$ and $g_B$ remains deterministic and study of closed form expression is important to find the optimal resource allocation strategy (i.e., how to choose $\epsilon$ and $P_J$). Considering $\mathbf{h}$ and $g_B$ as deterministic makes the study completely different from that in \cite{gaojie2017} as the Laplace trick used there can not be directly applied to derive the SOP closed form expression.

\subsection{Full-Duplex Bob in the TAS scheme}\label{TAS C FD}

The SOP against colluding EDs conditional on $\mathbf{h}$ and $g_B$ is
\begin{align} \small
&P[S_{AB}^{TAS}< R_s|\mathbf{h},g_B]=P\left[\frac{1+SNR_{AB}^{TAS}}{1+\sum\limits_{e\in\Phi}SNR_{AE_e}^{TAS}}
<2^{R_S}|\mathbf{h},g_B \right]\nonumber\\
&=P\left[\sum\limits_{e\in\Phi}\frac{|h_{A_{i^*}E_e}|^2}{\frac{d_{AE_e}^{\alpha}}{P_T}+
\frac{d_{AE_e}^{\alpha}}{d_{BE_e}^{\alpha}} mX_2}>y_0\right]\nonumber\\
&=\int_{y_0}^\infty f_{I_e}(x)\mbox{d}x
\end{align}
where
 $I_e=\sum\limits_{e\in\Phi}SNR_{AE_e}^{TAS}=\sum\limits_{e\in\Phi}\frac{|h_{A_{i^*}E_e}|^2}
 {\frac{d_{AE_e}^{\alpha}}{P_T}+\frac{d_{AE_e}
^{\alpha}}{d_{BE_e}^{\alpha}}mX_2}$ which is the sum of SNRs at all EDs. It is shown in Appendix \ref{sec:Laplace} that the Laplace transform of the PDF of $I_e$ is
\begin{eqnarray}\label{eq:laplace_FD_TAS} \small
\mathcal{L}_{I_e}(s)=\exp\bigg[-\rho_E\int_{0}^{R}\int_{0}^{2\pi}\frac{s}{f_e}\mathbf{E_1}(K(s))e^{K(s)}\mbox{d}\theta r\mbox{d}r\bigg]
\end{eqnarray}
where $\mathbf{E_1}(a)=\int_0^\infty\frac{e^{-ax}}{1+x}\mbox{d}x$ is the so called exponential integral function of $a$ and $K(s)=\frac{s+\frac{d_{AE_e}^{\alpha}}{P_T}}{f_e}$. Note that $\mathbf{E_1}(a)$ is monotonically decreasing function of $a$, and $K(s)$ is a strictly positive quantity. Later, we will discuss the relationship between $\mathbf{E_1}(K(s))$ and SOP.

We know that
\begin{align}\label{eq:SOP_laplace_FD_TAS} \small
&P[S_{AB}^{TAS}<R_s|\mathbf{h},g_B]=P[\frac{I_e}{y_0}>1]\nonumber\\
&\lessapprox P[\frac{I_e}{y_0}>l]\nonumber\\
&= \mathbb{E}\big[1-\exp(-\frac{aI_e}{y_0})\big]^N\nonumber\\
&= \mathbb{E}\left [\sum_{n=0}^{N}{{N}\choose{n}}(-1)^n\exp(-\frac{an}{y_0}I_e)\right ]\nonumber\\
&=\sum_{n=0}^{N}{{N}\choose{n}}(-1)^n\mathcal{L}_{I_e}
\big(\frac{an}{y_0}\big),
\end{align}
where $\lessapprox$ denotes ``less than and asymptotically equal to'', and $l$ is a normalized gamma distributed random variable with the shape parameter $N$, and as $N \to \infty$, $l$ approaches its upper bound equal to 1 \cite{Ju2018}-\cite{TBai2015}.  (Note that the left side of $\lessapprox$ is less than the right side if $N$ is finite, or equals to the right side if $N \to \infty$.)  Also $a=\frac{N}{(N!)^{\frac{1}{N}}}$, and $y_0$ is a realization of $Y$.

From \eqref{eq:laplace_FD_TAS} and \eqref{eq:SOP_laplace_FD_TAS}, we have
\begin{prop}\label{TAS_C}
For the TAS scheme,
\begin{align}\label{eq:SOP_colluding_FD_TAS} \small
&P[S_{AB}^{TAS}<R_s|\mathbf{h},g_B]\lessapprox\sum_{n=0}^{N}{{N}\choose{n}}(-1)^n\exp\bigg[-\rho_E\nonumber\\
&\qquad\int_{0}^{R}\int_{0}^{2\pi} \frac{s}{f_e}\mathbf{E_1}(K(s))e^{K(s)}\mbox{d}\theta r\mbox{d}r\bigg] \bigg|_{s=\frac{an}{y_0}}
\end{align}
where $s=\frac{an}{y_0}$, $K(s)=K(s)\big|_{s=\frac{an}{y_0}}=K(s)\big|_{s=\frac{an}{\frac{y}{\beta}-1+\frac{1}{\beta}}}=\frac{d_{BE_e}^{\alpha}}{P_J}+
\frac{s}{f_e}=\frac{d_{BE_e}^{\alpha} }{P_J}+\frac{an}{f_ey_0}$. If $\beta=1$, we have $y_0=y$ and then $K(s)=\frac{d_{BE_e}^{\alpha}}{P_J}+an\frac{d_{BE_e}^{\alpha}}{d_{AE_e}^{\alpha}}
\frac{\frac{1}{P_J}+\rho|g_B|^2}{\max\limits_{i\in M}|h_i|^2}$ which is independent of $P_T$. \\

\end{prop}

\begin{remark}
The secrecy performance is dependent on $P_J$ throughout the term $\Xi(s,r,\theta)|_{s=\frac{an}{y_0}}= \frac{s}{f_e}\mathbf{E_1}(K(s))e^{K(s)}|_{s=\frac{an}{y_0}}$. One can verify that as $P_J$ increases,
\begin{itemize}
	\item $\frac{s}{f_e}e^{K(s)}$ decreases monotonically and saturates to a lower bound.
	\item $\mathbf{E_1}(K(s))$ increases monotonically and saturates to an upper bound.
\end{itemize}
\end{remark}

These statements indicate that finding the optimal $P_J$ to minimize $\Xi(s,r,\theta)|_{s=\frac{an}{y_0}}$ is similar to that of $\Psi(Y,r,\theta)$ for the non-colluding TAS scheme. A comparison between $\Psi(Y,r,\theta)$ and $\Xi(s,r,\theta)|_{s=\frac{an}{y_0}}$ is shown in Table \ref{Xi table}. Note that, optimum jamming power ${P_J}^*$ is not necessarily the same for $\Psi(Y,r,\theta)$ and $\Xi(s,r,\theta)|_{s=\frac{an}{y_0}}$. Finally, simulation result shows that as $P_J$ increases, the conditional SOP in \eqref{eq:SOP_colluding_FD_TAS} achieves its minimum at a finite nonzero $P_J$.

\begin{table}[t]
\centering
\caption{Comparison between $\Psi(Y,r,\theta)$ and $\Xi(s,r,\theta)|_{s=\frac{an}{y_0}}$ subject to $R_s=0$. The column for $P_J \to \infty$ is subject to a fixed $\rho P_J$.}\label{Xi table}
\small\addtolength{\tabcolsep}{-6pt}
\begin{tabular}[t]{|c|c|c|c|c|}
\hline
   & $P_T$ & $P_J \to \infty$ & $P_J \uparrow (\leq {P_J }^*)$ &$P_J \uparrow (\geq {P_J }^*)$ \\
\hline
  $\Psi(Y,r,\theta)$  & $-$ & $\to 0$ & $\downarrow$ & $\uparrow$\\
\hline
  $\Xi(s,r,\theta)|_{s=\frac{an}{y_0}}$  & $-$ & $\to const.$ & $\downarrow$ & $\uparrow$\\

\hline
\end{tabular}
\end{table}%

\subsection{Half-Duplex Bob in TAS scheme} \label{TAS C HD}

If Bob is in the HD mode, then the sum of ED's SNR is $I_e^{HD}=\sum\limits_{e\in\Phi} \frac{X_{1,e}^{'}P_T}{d_{AE_e}^{\alpha}}$, where $X_{1,e}^{'}$ is exponentially distributed with unit mean. One can verify that the Laplace transform of the PDF of $I_e^{HD}$ is
\begin{align}\label{eq:37} \small
&\mathcal{L}_{I_e^{HD}}(s)=E_{\Phi}\bigg[\prod_{e\in\Phi}\frac{1}{1+\frac{sP_T}{d_{AE_e}^{\alpha}}}\bigg]\nonumber\\
&=\exp\bigg[-\rho_E\pi R^2\mathbf{F}(1,\frac{2}{\alpha};1+\frac{2}{\alpha};-\frac{R^{\alpha}}{sP_T})\bigg]
\end{align}
where $\mathbf{F}(1,\frac{2}{\alpha};1+\frac{2}{\alpha};-\frac{R^{\alpha}}{sP_T})$ is known as the Gaussian hypergeometric function. Note that $\alpha$ is governed by the environment. So, only the last parameter $\frac{R^{\alpha}}{sP_T}$ in $\mathbf{F}(1,\frac{2}{\alpha};1+\frac{2}{\alpha};-\frac{R^{\alpha}}{sP_T})$ is controllable via $P_T$, which takes real value between 0 to $\infty$. One can verify that $\frac{R^{\alpha}}{sP_T}$ is independent of $P_T$ for $\beta
=1$, which is similar as non-colluding HD TAS scheme.

Replacing $\mathcal{L}_{I_e}$ in \eqref{eq:SOP_laplace_FD_TAS} by  $\mathcal{L}_{I_e^{HD}}$ in \eqref{eq:37} yields
\begin{align}\label{eq:SOP_colluding_HD_TAS} \small
&P[S_{AB}^{TAS}<R_s|\mathbf{h},g_B]\lessapprox\sum_{n=0}^{N}{{N}\choose{n}}(-1)^n\exp\bigg[-\rho_E\pi R^2\nonumber\\
&\qquad \mathbf{F}(1,\frac{2}{\alpha};1+\frac{2}{\alpha};-\frac{R^{\alpha}}{sP_T})\bigg]
\bigg|_{s=\frac{an}{y_0}}
\end{align}
where $s=\frac{an}{y_0}$. The result in \eqref{eq:SOP_colluding_HD_TAS} is that in \eqref{eq:SOP_colluding_FD_TAS} with $P_J=0$ but the former is a much simplified form than the latter.

\subsection{Full-Duplex Bob in TAB scheme without AN from Alice} \label{TAB C FD}
Conditional on $\mathbf{h}$ and $g_B$, the legitimate channel's SNR is $z$ as previously defined. For $\epsilon=0$ (i.e., without AN from Alice), the SNR at the $e$th Eve  is (from \eqref{eq:SNR_AE_TAB}):
\begin{align} \small
SNR_{AE_e}^{TAB}&=\frac{X_1\Theta}{\frac{d_{AE_e}^\alpha}{P_T}+m\frac{d_{AE_e}^\alpha}{d_{BE_e}^\alpha}X_2}\nonumber\\
&=\frac{X_4}{\frac{d_{AE_e}^\alpha}{P_T}+m\frac{d_{AE_e}^\alpha}{d_{BE_e}^\alpha}X_2}.
\end{align}
Similar to the analysis leading to \eqref{eq:SOP_colluding_FD_TAS}, the SOP now is still given by \eqref{eq:SOP_colluding_FD_TAS}  but with  $s=\frac{an}{\frac{z}{\beta}-1+\frac{1}{\beta}}$. Hence, we have:

\begin{prop}\label{TAB_C}
For the TAB scheme with $\epsilon=0$,
\begin{align}\label{eq:SOP_colluding_FD_TAB}
&P[S_{AB}^{TAB}<R_s|\mathbf{h},g_B]\lessapprox\sum_{n=0}^{N}{{N}\choose{n}}(-1)^n\exp\bigg[-\rho_E\nonumber\\
&\qquad\int_{0}^{R}\int_{0}^{2\pi} \frac{s}{f_e}\mathbf{E_1}(K(s))e^{K(s)}\mbox{d}\theta r\mbox{d}r\bigg] \bigg|_{s=\frac{an}{\frac{z}{\beta}-1+\frac{1}{\beta}}
}.
\end{align}
\end{prop}

 Since $\|\mathbf{h}\|^2\geq\max\limits_{i\in M}|h_i|^2$, the TAB scheme always outperforms the TAS scheme.

 \subsubsection{Bob in Half-Duplex Mode} \label{TAB C HD}
In this case, we have $P_J=0$ and $z=P_T \|\mathbf{h}\|^2$. The SOP expression is similar to \eqref{eq:SOP_colluding_HD_TAS} and can be expressed as
\begin{align}\label{eq:SOP_colluding_HD_TAB} \small
&P[S_{AB}^{TAB}<R_s|\mathbf{h},g_B]\lessapprox\sum_{n=0}^{N}{{N}\choose{n}}(-1)^n\exp\bigg[-\rho_E\pi R^2\nonumber\\
&\qquad \mathbf{F}(1,\frac{2}{\alpha};1+\frac{2}{\alpha};-\frac{R^{\alpha}}{sP_T})\bigg]
\bigg|_{s=\frac{an}{\frac{z}{\beta}-1+\frac{1}{\beta}}}.
\end{align}

\subsection{Full-Duplex Bob in TAB scheme with AN from Alice} \label{TAB C E}
For the TAB scheme with $\epsilon >0$, we have
\begin{eqnarray} \small
&&\!\!\!\!\!\!\!\!P[S_{AB}^{TAB}<R_s|\mathbf{h},g_B]=P[\frac{1+SNR_{AB}^{TAB}}{1+\sum\limits_{e\in\Phi}SNR_{AE_e}^{TAB}}<2^{R_S}|\mathbf{h},g_B]\nonumber\\
&&\overset{(a)}\lessapprox P[\sum\limits_{e\in\Phi}\frac{X_4}{\frac{\epsilon}{M-1}X_{4,4}+f_eX_2}>\frac{z}{\beta}+
\frac{\frac{1}{\beta}-1}{1-\epsilon}]\nonumber\\
&&\overset{(b)}\lessapprox\sum_{n=0}^{N}{{N}\choose{n}}(-1)^n\mathcal{L}_{\tilde{I}_e}\big(\frac{an}{\frac{z}{\beta}+\frac{\frac{1}{\beta}-1}{1-\epsilon}}\big)
\end{eqnarray}
where the parameters defined after (\ref{UE_order}) have been applied and $\tilde{I}_e=\sum\limits_{e\in\Phi} \frac{X_4}{\frac{\epsilon}{M-1}X_{4,4}+f_eX_2}$. Here, $(a)$ is due to neglecting the background noise $n_{A,E_e}(k)$ at Eve (but not the noise at Bob), and $(b)$ is due to the application of the normalized gamma random variable as discussed before. Similar to that in Appendix \ref{sec:Laplace}, one can verify
\begin{align}\label{eq:laplace_FD_TAB} \small
&\mathcal{L}_{\tilde{I}_e}(s)=\exp\bigg[-\rho_Es\int_{0}^{R}\int_{0}^{2\pi}\int_{x=0}^{\infty}\frac{e^{-sx}}{(1+f_e x)}\nonumber\\
&\times \frac{1}{(1+\frac{\epsilon}{M-1}x)^{M-1}}\mbox{d}x\mbox{d}\theta r\mbox{d}r\bigg].
\end{align}

\section{Simulations}\label{simulation}

In this section, we illustrate the secrecy outage probabilities (SOP) of the TAS and TAB schemes against randomly located EDs. We consider both colluding and non-colluding cases. Most of our simulation results provide comparisons between TAS and TAB schemes. Moreover, we present the secrecy performance enhancement of the TAB scheme using AN.

Throughout the simulations, we will assume unit noise variance, $\alpha=2$, $P_T=40$ dB, $R_D=4$ b/s/Hz, $\rho_E=1$, $M=5$, $d=1$ and $R=5$. Unless otherwise specified, we let $P_J$, $\rho$ and $\epsilon$ be $40$ dB, $0.01$ and $0.01$ respectively. Since Alice can estimate the legitimate channel and know the self interference channel of Bob, therefore, we will first study the SOP under conditional $\mathbf{h}$ and $g_B$ for the TAB scheme. Considering $R_D=4$, $\epsilon$ can be set between $0$ and $0.53$ to maintain a nonzero desired data transmission for the above given $\rho$ and $P_J$.

\begin{figure}[ht]
	\vspace{-0.1in}
	\begin{center}
		\includegraphics[width = .45\textwidth, height=.3\textwidth]{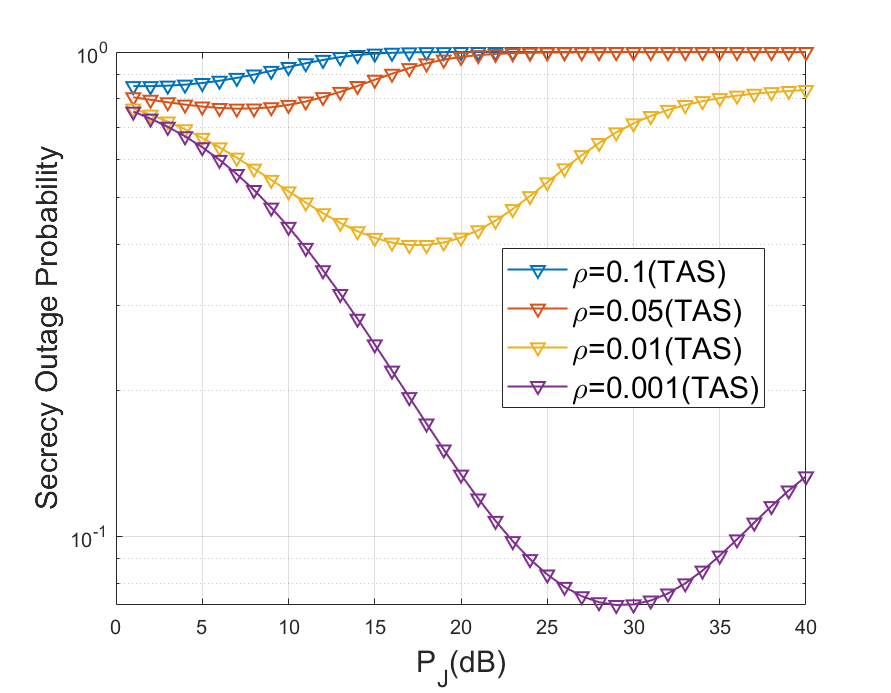}
		\vspace{-0.1in}
		\includegraphics[width = .45\textwidth, height=.3\textwidth]{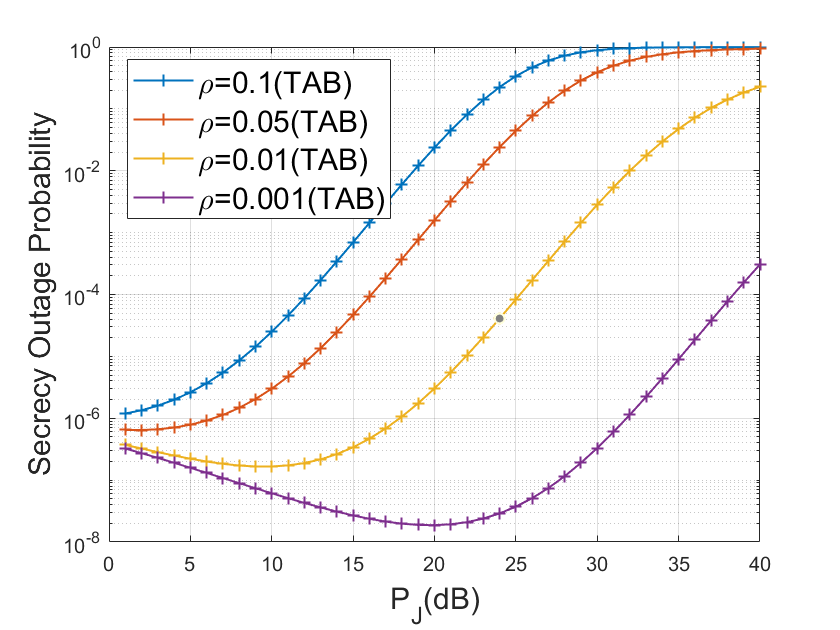}
		\caption{Comparison of the TAS and TAB schemes in terms of $P_{out}$ against non-colluding EDs.}\label{optimum Pj NC}
	\end{center}
\end{figure}

In Fig. \ref{optimum Pj NC}, the SOP of the TAS and TAB schemes for non-colluding EDs is illustrated under different values of $P_J$ and $\rho$. For the TAB scheme, $\epsilon=0$ is chosen. We see that as $\rho$ decreases, the optimum jamming power increases which results in lower SOP for both TAS and TAB schemes. And the TAB scheme outperforms the TAS scheme substantially.

\begin{figure}[ht]
\vspace{-0.1in}
	\begin{center}
		\includegraphics[width = .45\textwidth, height=.3\textwidth]{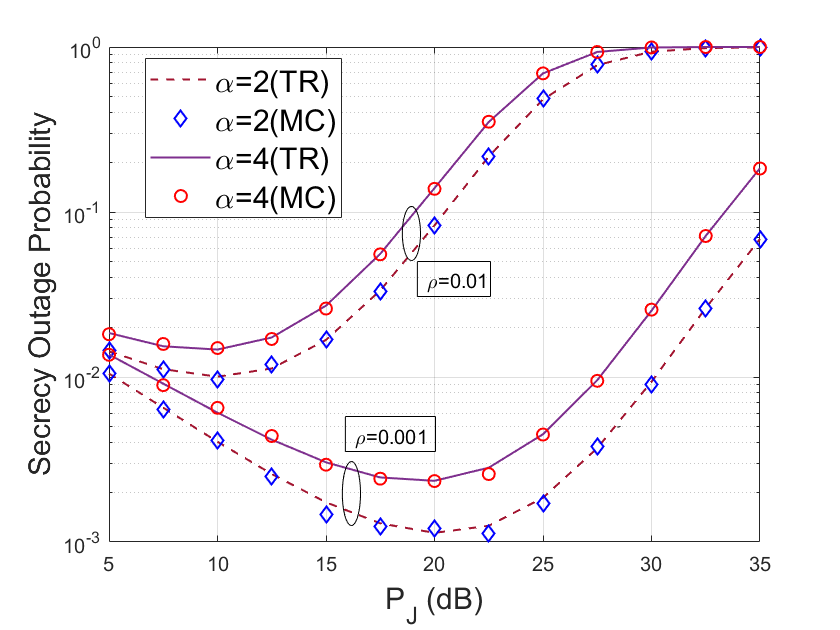}
		\caption{Comparison of theoretical results (``TR'') and simulation results (``MC'') of the TAB scheme in terms of $P_{out}$ versus $P_J$.}\label{optimum Pj NC MC}
	\end{center}
\end{figure}

In Fig. \ref{optimum Pj NC MC}, we compare the Monte Carlo (MC) simulation results (using $N_R=10^5$ independent runs) with our theoretical results shown in \eqref{eq:TAB} where $R=5$ and $\rho_E=10$.
We observe that the two results match each other very well. This consistency between theory and simulation holds for all other results we have tested under a sufficiently large $N_R$.

\begin{figure}[ht]
\vspace{-0.1in}
	\begin{center}
		\includegraphics[width = .45\textwidth, height=.3\textwidth]{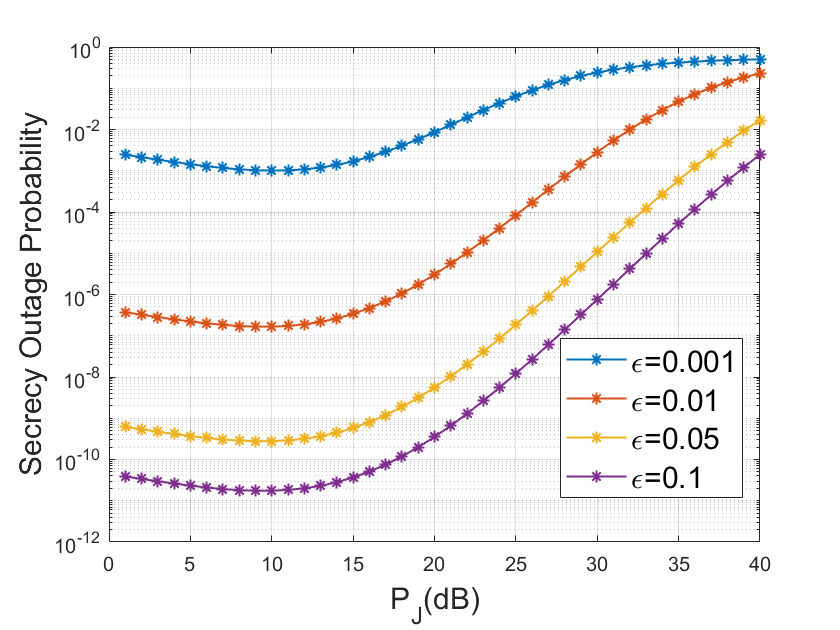}
		\caption{Illustration of $P_{out}=1-P_{con}$ versus $P_J$ and $\epsilon$ for the TAB scheme.}
		\label{fig:eps_varying}
	\end{center}
\end{figure}

Fig. \ref{fig:eps_varying} shows the SOP of the TAB scheme with $\epsilon>0$. We see that the SOP decreases as $\epsilon$ increases, the optimal value of $P_J$ is dependent on $\epsilon$ but the dependence is rather weak (or not very sensitive).

To illustrate the TAS and TAB schemes with user selection (i.e., TAS-US and TAB-US), we consider $P_T=50$ dB, $\alpha=2$, $\beta=2$, $\epsilon=0.00001$, $\rho_U =0.5$ and $\rho_E=0.1$ unless otherwise specified.

\begin{figure}[ht]
     \vspace{-0.1in}
     \begin{center}
		\includegraphics[width = .45\textwidth, height=.3\textwidth]{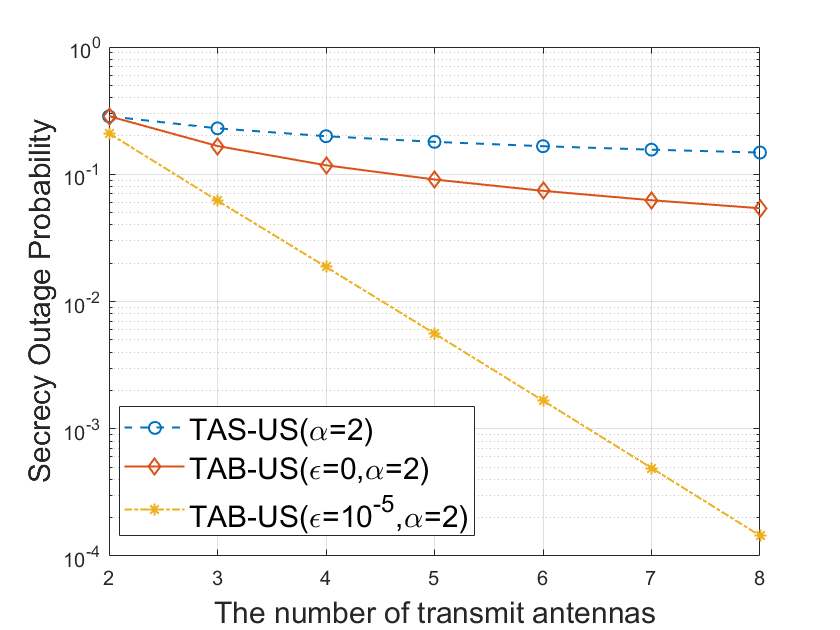}
		\caption{SOP of TAS-US and TAB-US for the nearest user vs the number of transmit antennas  against non-colluding EDs.}
		\label{SOP_vs_M}
	\end{center}
\end{figure}

Fig. \ref{SOP_vs_M} shows the SOP of the TAS-US and TAB schemes for the nearest user. As the number $M$ of transmit antennas increases, the performance gap between TAB-US and TAS-US increases rapidly for $\epsilon>0$. More importantly, we see that only a small fraction (e.g., $\epsilon=0.00001$ or $\epsilon P_T=0$ dB which is at the same level as the noise variance) of the transmit power used for AN makes a huge difference.

\begin{figure}[h]
	\vspace{-0.1in}
	\begin{center}
		\includegraphics[width = .45\textwidth, height=.3\textwidth]{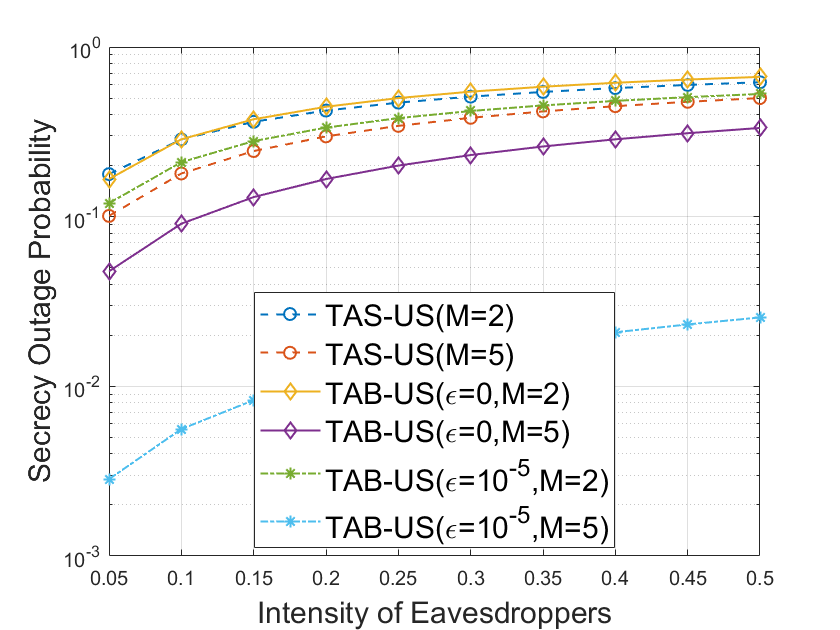}
		\caption{SOP vs intensity of eavesdroppers for ordered users against non-colluding EDs.}
		\label{SOP_vs_rho_E}
	\end{center}
\end{figure}

\begin{figure}[h]
     \vspace{-0.1in}
     \begin{center}
		\includegraphics[width = .45\textwidth, height=.3\textwidth]{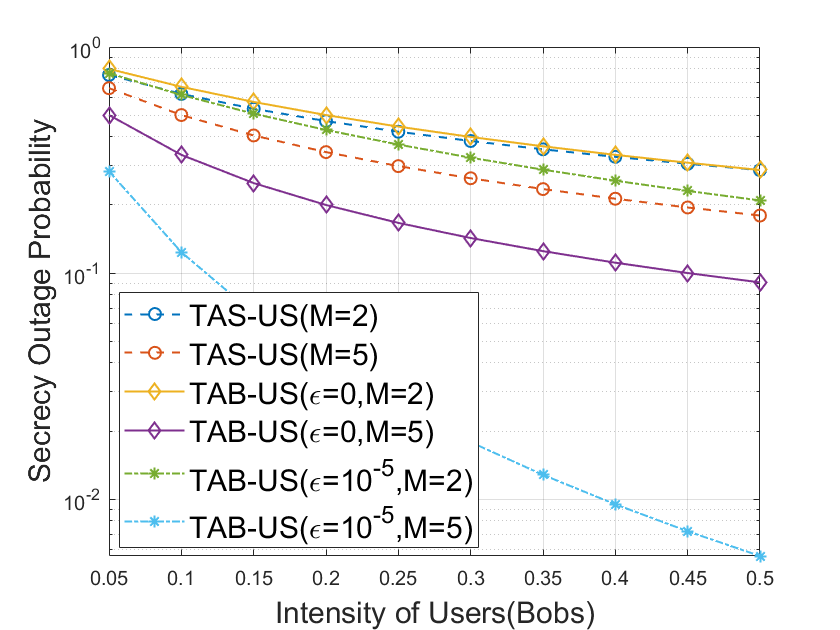}
		\caption{SOP vs intensity of users against non-colluding EDs.}
		\label{SOP_vs_rho_U}
	\end{center}
\end{figure}

Fig. \ref{SOP_vs_rho_E} illustrates the effects of ED's density $\rho_E$ on the SOP of TAS-US and TAB-US for the nearest user. And Fig. \ref{SOP_vs_rho_U} illustrates the effects of the users' density $\rho_U$ on the SOP of TAS-US and TAB-US for the nearest user. We see that SOP increases as $\rho_E$ increases but decreases as $\rho_U$ increases. The performance gap between TAS-US and TAB-US remains approximately the same as $\rho_E$ increases but increases as $\rho_U$ increases.

\begin{figure}[h]
     \vspace{-0.1in}
	\begin{center}
		\includegraphics[width = .45\textwidth, height=.3\textwidth]{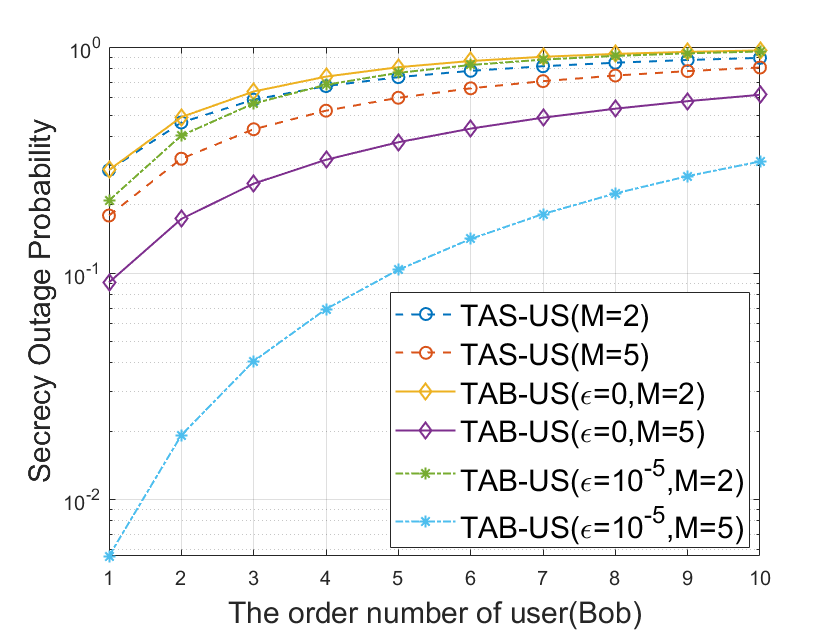}
		
		\caption{SOP vs the order number of user against non-colluding EDs.}
		\label{SOP_vs_n}
	\end{center}
\end{figure}

Fig. \ref{SOP_vs_n} shows the SOP of the TAS-US and TAB-US schemes as functions of the order index ($n$) of users (from nearest to farthest). We see that the SOP increases as $n$ increases and the performance gap between TAB-US and TAS-US reduces as $n$ increases.

\begin{figure}[h]
	\vspace{-0.1in}
	\begin{center}
		\includegraphics[width = .45\textwidth, height=.3\textwidth]{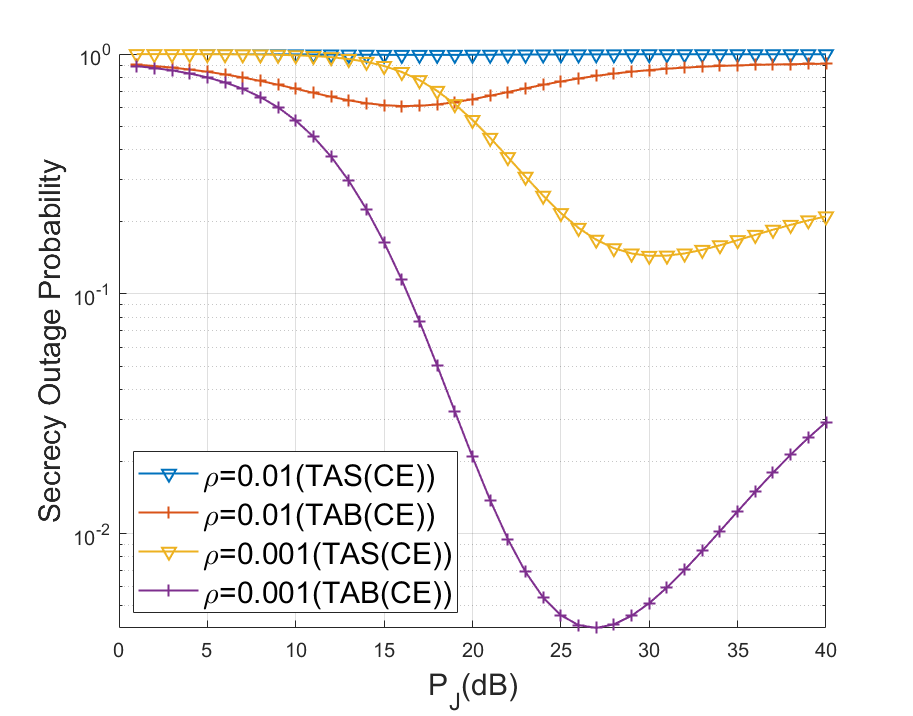}
		\caption{Comparison of the TAB and TAS schemes in terms of $P_{out}$ against colluding EDs.}
		{\label{colluding_Pj}}
	\end{center}
\end{figure}

Now, we consider the TAB and TAS schemes for colluding EDs. We assume that there are two circles of radii $R_g$ and $R$ around Alice, and EDs exist and collude within the two circles. In our experiment, we let $R_g=0.1$ and $R=5$.
Although the closed form expressions of the SOP in this case are all in series expansions, choosing $N=20$ (e.g., see \eqref{eq:SOP_colluding_FD_TAS}) provided good approximations.

Fig. \ref{colluding_Pj} shows the SOP of TAS and TAB schemes for colluding EDs as functions of  the self-interference power gain $\rho$ and the jamming power $P_J$ from full-duplex Bob. We see that the optimal $P_J$ increases as the self-interference power gain $\rho$ decreases, and the optimized SOP reduces significantly as $\rho$ decreases.

\begin{figure}[h]
	\vspace{-0.1in}
	\begin{center}
		\includegraphics[width = .45\textwidth, height=.3\textwidth]{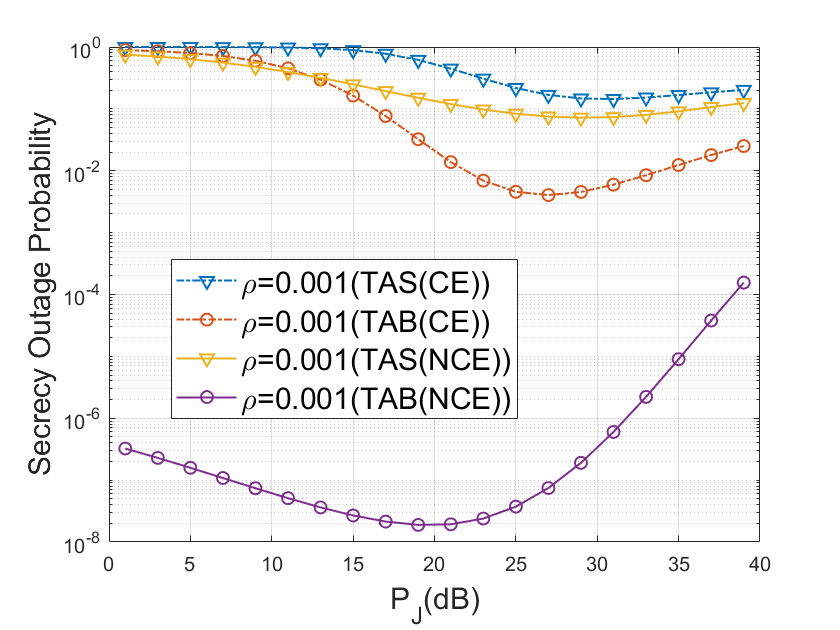}
		\caption{Comparison of the TAB and TAS schemes in terms of $P_{out}$ against colluding and non-colluding EDs.}
		\label{final_analysis}
	\end{center}
\end{figure}

Finally, Fig. \ref{final_analysis} illustrates the differences of SOP for colluding and non-colluding EDs. We see that the performance gap between colluding and non-colluding is large. But the TAB scheme is consistently better than the TAS scheme in terms of SOP.
\vspace{-.1in}
\section{Conclusion}

In this paper, we presented closed form expressions of secrecy outage probabilities (SOP) of several schemes for multi-antenna downlink transmissions against randomly located eavesdroppers (EDs). We considered both transmit antenna selection (TAS) and transmit antenna beamforming (TAB) schemes, full-duplex (FD) and half-duplex (HD) receivers/users, colluding and non-colluding EDs, the use of artificial noise (AN) from transmitter, and user selection based on their distances to the transmitter. For all these schemes and scenarios, we assume that EDs are distributed as the Poisson Point Process (PPP). For user selection, we also assume the PPP model for users' locations. The closed-form expressions of SOP are useful for numerical computations needed for network design purposes. We provided numerical examples to illustrate the usefulness of these expressions, which also revealed important observations such as the optimal jamming power from FD users and the impacts of several other parameters on SOP.

\appendices

\section{Proof of \eqref{eq:Piny}}\label{sec:TAS}
It follows from \eqref{eq:SAB}, \eqref{eq:SNR_AB_TAS} and \eqref{eq:SNR_AE_TAS} that
\begin{eqnarray} \small
P_{con,\Phi,Y}&\overset{\Delta}{=}&P[S_{AB}^{TAS}>R_s|\Phi,Y]\nonumber \\
&=&P\left[\left.\frac{1+SNR_{AB}^{TAS}}{1+\max\limits_{e \in\Phi}SNR_{AE_e}^{TAS}}>2^{R_S}\right|\Phi,Y\right]\nonumber\\
&=&P\left[\left.\max\limits_{e\in\Phi}SNR_{AE_e}^{TAS}<Y_0\right|\Phi,Y\right]\nonumber\\
&=&\prod\limits_{e\in\Phi}P\left[\left.\frac{|h_{A_{i^*}E_e}|^2}{\frac{d_{AE_e}^{\alpha}}{P_T}+\frac{d_{AE_e}^{\alpha}}{d_{BE_e}^{\alpha}}m|h_{BE_e}|^2}<Y_0\right|\Phi,Y \right]\nonumber \\
&=&\prod\limits_{e\in\Phi}\big(1-\Psi(Y,r_e,\theta_e)\big) ,\label{eq:PinT}
\end{eqnarray}
where (due to the lemma shown next)
\begin{equation}\label{introduce_psi}\small
\Psi(Y,r_e,\theta_e)=\frac{\exp(-\frac{{d_{AE_e}^{\alpha}}}{P_T}Y_0)}{1+m(\frac{d_{AE_e}}{d_{BE_e}})^{\alpha}Y_0} \nonumber
\end{equation}
 We have applied the following lemma.
\begin{lemma}
	If $A$ and $B$ (like $|h_{A_{i^*}E_e}|^2$ and $|h_{BE_e}|^2$ ) are two independent random variables with the exponential distribution of unit mean, then $P(\frac{A}{a+bB}<c)=1-\frac{e^{-ac}}{1+bc}$.
\end{lemma}

Note that we are only interested in such $R_s$ that $\log_2(1+SNR_{AB})\geq R_s$, which implies $Y_0\geq0$.

Let $P_{con,Y}$ be $P_{con}$ conditional only on $Y$. Applying the Campbell's theorem \cite{campbell} to \eqref{eq:PinT} yields:
\begin{eqnarray} \small
P_{con,Y}&=&\mathbb{E}_{\Phi}\{P[SNR_{AB}^{TAS}>R_s|\Phi,Y]\}\nonumber\\
&=&\exp\bigg[-\rho_E\int_{0}^{R}\int_{0}^{2\pi}\Psi(Y,r,\theta)r\mbox{d}\theta \mbox{d}r\bigg]. \nonumber
\end{eqnarray}
\section{A simplification of the double integral in \eqref{eq:Piny}}\label{sec:simplification}
Assume $R_s=0$ and $\alpha=2$. Then, $\beta=1$ and $Y_0=Y$. Let the distance between Alice and Bob be $d$. Then, $\frac{d_{AE}^\alpha}{d_{BE}^\alpha}=\frac{d_{AE}^2}{d_{BE}^2}=\frac{r^2}{r^2+d^2-2rd\cos\theta}$, and furthermore
\begin{eqnarray} \small
&&\int_{0}^{R}\int_{0}^{2\pi}\Psi(Y,r,\theta)r\mbox{d}\theta\mbox{d}r\nonumber \\
&=&\int_{0}^{R}\int_{0}^{2\pi}\frac{\exp(-\frac{Yr^2}{P_T})}{1+mY\frac{r^2}{r^2+d^2-2rd\cos\theta}}r\mbox{d}\theta \mbox{d}r\nonumber\\
&=&\int_{0}^{R}\int_{0}^{2\pi} \exp(-\frac{Yr^2}{P_T})\nonumber\\
&&\times\frac{\big((1+mY)r^2+d^2-2rd\cos\theta\big)-mYr^2}{(1+mY)r^2+d^2-2rd\cos\theta} r\mbox{d}\theta\mbox{d}r \nonumber\\
&=&2\pi\int_{0}^{R}\exp(-\frac{Yr^2}{P_T})r\mbox{d}r\nonumber\\
&&-\int_0^R  \int_0^{2\pi}\frac{mYr^3\exp(-\frac{Yr^2}{P_T})}{(1+mY)r^2+d^2-2rd
\cos\theta} \mbox{d}\theta \mbox{d}r,\label{eq:two_terms}
\end{eqnarray}

where the first term can be obviously reduced. The second term in \eqref{eq:two_terms}, can be simplified by applying $\int_0^{2\pi}\frac{1}{a-b\cos\theta}\mbox{d}\theta=\frac{2\pi}{\sqrt{a^2-b^2}}$ identity. Then, \eqref{eq:two_terms} yields
\begin{eqnarray}\label{eq:simplification} \small
&&\int_{0}^{R}\int_{0}^{2\pi}\Psi(Y,r,\theta)r\mbox{d}\theta \mbox{d}r\nonumber\\
&=&\frac{\pi P_T}{Y}\left(1-\exp\left(-\frac{YR^2}{P_T}\right)\right)-2\pi mY\nonumber\\
&\quad& \times \int_{0}^{R}\frac{r^3\exp(-\frac{Yr^2}{P_T})}{\sqrt{\big((1+mY)r^2+d^2\big)^2-4r^2d^2}}\mbox{d}r\nonumber\\
&=&\frac{\pi P_T}{Y}\bigg(1-\exp\left(-\frac{YR^2}{P_T}\right)\bigg)-\pi mY\nonumber\\
&\quad& \times\int_{0}^{R^2}\frac{\exp\left(-\frac{Yr}{P_T}\right)r}{\sqrt{\big((1+mY)r+d^2\big)^2-4rd^2}}\mbox{d}r.
\end{eqnarray}
which is a much simplified expression of the double integral in \eqref{eq:Piny}.
\section{Proof of Lemma \ref{lemma2}} \label{lemma2proof}
Here, $Z=\frac{\|\mathbf{h}\|^2}{ \frac{1}{P_T}+\rho m|g_B|^2}$. Lets consider, $Y_3=\|\mathbf{h}\|^2$, $Y_2=\frac{1}{P_T}+\rho m|g_B|^2$ and $Z=\frac{Y_3}{Y_2}$. Note that $f_{Y_3}(x)=\frac{1}{\Gamma(M)}x^{M-1}e^{-x}$ and $f_{Y_2}(x)= \frac{1}{\rho m}exp(-\frac{x-\frac{1}{P_T}}{\rho m}), x>\frac{1}{P_T}$.
\vspace{-.05in}
\begin{eqnarray} \small
F_Z(z) &=& P[\frac{Y_3}{Y_1} < z] \nonumber \\
&=& P[ |g_B|^2 > \frac{Y_3- \frac{z}{P_T}}{z\rho m} ] \nonumber \\
&=& \int_{y=0}^{\infty} f_{Y_3}(y)dy  \int_{x=\frac{y- \frac{z}{P_T}}{z\rho m}}^{\infty} e^{-x}dx \nonumber \\
&=& \frac{1}{\Gamma(M)}\int_{y=0}^{\infty} y^{M-1}e^{-y}e^{-\frac{y- \frac{z}{P_T}}{z\rho m}} dy \nonumber \\
&=& \frac{e^{\frac{1}{\rho  P_J}}}{\Gamma(M) (1+\frac{1}{z\rho m})^{M}}\int_{y=0}^{\infty} {\big(y(1+\frac{1}{z\rho m})}\big)^{M-1} \nonumber \\
&& \times e^{-y(1+\frac{1}{z\rho m})} d\big( y(1+\frac{1}{z\rho m})\big) \nonumber \\
&=& \frac{e^{\frac{1}{\rho  P_J}}}{ (1+\frac{1}{z\rho m})^{M}}.
\end{eqnarray}
If $m>0$ then, $f_Z(z)={e^{\frac{1}{\rho  P_J}} M\rho m}\frac{{z\rho m}^{M-1}}{ (1+{z\rho m})^{M+1}}$. Ror $m=0$, $Z$ follows scaled CHI squared distribution.
\section{Proof of \eqref{eq:TAB}}\label{sec:TAB}
The secrecy of the TAB scheme can be analyzed as follows.
\begin{eqnarray}\label{TAB_perf} \small
&&P_{con,\Phi,\mathbf{h},g_B}\overset{\Delta}{=} P[S_{AB}>R_S|\Phi,\mathbf{h},g_B] \nonumber\\
&=& P[S_{AB}>R_S|\Phi,z] \nonumber\\
&=&P\left[\left.\max\limits_{e \in\Phi} SNR_{AE_e}^{TAB}<\frac{SNR_{AB}}{\beta}-(1-\frac{1}{\beta})\right|\Phi,z\right]\nonumber\\
&=&P\bigg[\max\limits_{e \in \Phi}\frac{X_1\Theta}{d_{AE}^\alpha+\frac{P_Jd_{AE_e}^\alpha}{d_{BE_e}^\alpha}X_2+\frac{\epsilon P_T}{M-1}X_1(1-\Theta)}\nonumber \\
&&\;\;\;\;\;\;\;\;\;\;\;\;<\left.\frac{{\|\mathbf{h}\|^2}}{\beta(1+\rho|g_B|^2P_J)}-\frac{1-\frac{1}{\beta}}
{(1-\epsilon)P_T}\right|\Phi,z\bigg]\nonumber\\
&=&\prod_{e \in\Phi}P\bigg[\bigg.\frac{X_1\Theta}{\frac{P_Jd_{AE_e}^\alpha}{d_{BE_e}^\alpha}\left(X_2+\frac{d_{BE}^\alpha}
{P_J}\right)+\frac{\epsilon P_T}{M-1} X_1(1-\Theta)}\nonumber\\
&&\;\;\;\;\;\;\;\;\;\;\;\;\;\;\;\;\;\;\;\;\;\;\;\;<{c}\bigg|\Phi,z \bigg] \nonumber\\
&=& \prod_{e \in\Phi}P\left[\left.{\Theta}<\frac{\frac{\epsilon}{M-1}+f_e\frac{X_3}{X_1}}{\frac{\epsilon}{M-1}+{g}}
\right|
\Phi,z\right ],
\end{eqnarray}
where $X_3=X_2+\frac{d_{BE}^\alpha}{P_J}$, and $X_1$, $X_2$ and $\boldsymbol{\Theta}$ are independent variables as defined previously (after \eqref{eq:SNR_AE_TAB}). It is easy to verify that
 $f_{\frac{X_2}{X_1}}(x)=M (1+x)^{-(M+1)}$, which is similar to the $F(2,2M)$ distribution
  \cite{Fdist}. When large scale channel gain of jamming signal at ED is higher than noise level, i.e., $\frac{P_J}{d_{BE}^\alpha}\gg 1$, the shift between $X_3$ and $X_2$ becomes smaller. Furthermore,  one can verify that $f_{\frac{X_3}{X_1}}(x)\approx e^{\frac{d_{BE}^\alpha}{P_J}}M (1+x)^{-(M+1)}$. It follows from the PDF $f_\Theta(x)$ shown earlier that the CDF of $\Theta$ is $F_\Theta(x)=1-(1-x)^{M-1}$. Then, it is shown in Appendix \ref{sec:P_Theta} that
\begin{equation}\label{eq:P_Theta}\small
P\left[{\Theta}>\frac{\frac{\epsilon}{M-1}+f_e\frac{X_3}{X_1}}{\frac{\epsilon}{M-1}+{g}}|\Phi,z\right]=
\frac{e^{\frac{d_{BE}^\alpha}{P_J}}}{(1+\frac{f_e}{{g}})(1+\frac{\epsilon}{(M-1){g}} )^{M-1}},
\end{equation}
where $g$ is a function of $Z$ as defined before.
Averaging over the PPP distribution of the locations of the Eves, one can verify (using the Campbell's theorem) that
\begin{align}\small
&P_{con,\mathbf{h},g_B}=P_{con,z} \overset{\Delta}{=}\mathbb{E}_{\Phi}\{P[S_{AB}>R_S|\Phi,\mathbf{h},g_B]\}\nonumber\\
&=\mathbb{E}_{\Phi}\left[\prod\limits_{e\in\Phi}\bigg(1-P\left[{\Theta}>\frac{\frac{\epsilon}{M-1}+
f_e\frac{X_3}{X_1}}{\frac{\epsilon}{M-1}+{g}}|\Phi,z\right]\bigg)\right]\nonumber\\
&=\mathbb{E}_{\Phi}\left[\prod\limits_{e\in\Phi}\bigg(1-\frac{e^{\frac{d_{BE}^\alpha}{P_J}}}
{(1+\frac{f_e}{{g}})(1+\frac{\epsilon}{(M-1){g}})^{M-1}}\bigg)\right]\nonumber\\
&=\exp\bigg[-\rho_E\int_{0}^{R}r\int_{0}^{2\pi}\Omega(\frac{1}{{g}};r,\theta)\mbox{d}\theta \mbox{d}r\bigg], \nonumber
\end{align}
where $z$ is a realization of $Z$, $g$ is a function of $z$, and
\begin{equation} \small
\Omega(\frac{1}{{g}};r,\theta)=\frac{e^{\frac{d_{BE}^\alpha}{P_J}}}{(1+\frac{f_e}{{g}})(1+\frac{\epsilon}
{(M-1){g}} )^{M-1}}. \nonumber
\end{equation}
\subsection{Proof of \eqref{eq:P_Theta}} \label{sec:P_Theta}
The complement of \eqref{eq:P_Theta} is
\begin{eqnarray}\label{appB} \small
&&P[{\Theta}<\frac{\frac{\epsilon}{M-1}+f_e\frac{X_3}{X_1}}{\frac{\epsilon}{M-1}+{g}}|\Phi,z]
\nonumber\\
&=&\int_{0}^{\infty}F_\Theta\left(\frac{\frac{\epsilon}{M-1}+f_ex}{\frac{\epsilon}{M-1}+{g}}\right)
f_{\frac{X_3}{X_1}}(x)\mbox{d}x\nonumber\\
&=&\int_{0}^{\frac{{g}}{f_e}}F_\Theta\left(\frac{\frac{\epsilon}{M-1}+f_ex}{\frac{\epsilon}{M-1}+{g}}\right)
f_{\frac{X_3}{X_1}}(x)\mbox{d}x\nonumber\\
&&\;\;\;\;\;\;\;\;+\int_{\frac{{g}}{f_e}}^\infty F_\Theta\left(\frac{\frac{\epsilon}{M-1}+f_ex}{\frac{\epsilon}{M-1}+{g}}\right)f_{\frac{X_3}{X_1}}(x)\mbox{d}x.
\end{eqnarray}

Here, $F_\Theta(y)=1$ for $y\geq1$, so $F_\Theta\left(\frac{\frac{\epsilon}{M-1}+f_ex}{\frac{\epsilon}{M-1}+{g}}\right)=1$ for $x\geq\frac{g}{f_e}$. Then (\ref{appB}) continues as follows:
\begin{align} \small
&\int_{0}^{\frac{{g}}{f_e}}F_\Theta\left(\frac{\frac{\epsilon}{M-1}+f_ex}{\frac{\epsilon}{M-1}+{g}}\right)
f_{\frac{X_3}{X_1}}(x)\mbox{d}x+\int_{\frac{{g}}{f_e}}^\infty f_{\frac{X_3}{X_1}}(x)\mbox{d}x\nonumber\\
&=\int_{0}^{\frac{{g}}{f_e}}\left[1-\left(1-\frac{\frac{\epsilon}{M-1}+f_ex}{\frac{\epsilon}{M-1}+{g}}\right)^{M-1}\right]f_{\frac{X_3}{X_1}}(x)\mbox{d}x\nonumber\\
&\qquad+1-\int_{x=0}^{\frac{{g}}
{f_e}}f_{\frac{X_3}{X_1}}(x)\mbox{d}x\nonumber\\
&=1-\frac{Me^{\frac{d_{BE_e}^\alpha}{P_J}}{g}^{M-1}}{(\frac{\epsilon}{M-1}+{g})^{M-1}}
\int_{0}^{\frac{{g}}{f_e}}\bigg(\frac{1-\frac{f_e}{{g}}x}{1+x}\bigg)^{M-1}\frac{\mbox{d}x}{(1+x)^2}.
\end{align}
Let $k=\frac{f_e}{{g}}$ and $z=\frac{1}{1+x}$. Then $(\frac{1-kx}{1+x})^{M-1}=k^{M-1}\big(-1+z\frac{k+1}{k}\big)^{M-1}$. The above leads to
\begin{align} \small
&P[{\Theta}<\frac{\frac{\epsilon}{M-1}+f_e\frac{X_3}{X_1}}{\frac{\epsilon}{M-1}+{g}}|\Phi,z]
\nonumber\\
&=1-\frac{Me^{\frac{d_{BE_e}^\alpha}{P_J}}({g}k)^{M-1}}{(\frac{\epsilon}{M-1}+{g})^{M-1}}
\int_{\frac{k}{k+1}}^{1}\big(-1+z(\frac{k+1}{k})\big)^{M-1}\mbox{d}z.
\end{align}
Now, using $y=z(\frac{k+1}{k})-1$, we have
\begin{eqnarray} \small
&&P[{\Theta}<\frac{\frac{\epsilon}{M-1}+f_e\frac{X_3}{X_1}}{\frac{\epsilon}{M-1}+{g}}|\Phi,\mathbf{h},g_B]\nonumber\\
&=&1-\frac{Me^{\frac{d_{BE_e}^\alpha}{P_J}}k^{M}}{(1+k)(1+\frac{\epsilon}{(M-1){g}})^{M-1}}\int_{y=0}^{y=
\frac{1}{k}}y^{M-1}\mbox{d}y\nonumber\\
&=&1-\frac{e^{\frac{d_{BE_e}^\alpha}{P_J}}}{(1+\frac{f_e}{{g}})(1+\frac{\epsilon}{(M-1){g}})^{M-1}}.
\end{eqnarray}

\section{Unimodality of $\Omega$}\label{proof of convexity}
From \eqref{OMEGA}, we have
\begin{equation} \small
    \Omega(\frac{1}{g};r,\theta)=\underbrace{\frac{e^{\frac{d_{BE_e}^\alpha}{P_J}}}{\left(1+\frac{f_e}{g}
    \right)}}_\textrm{${\Omega}_1(P_J)$}\underbrace{ \frac{1}{\left(1+\frac{\epsilon}{(M-1)g}\right)^{M-1}}}_\textrm{${\Omega}_2(P_J)$},
\end{equation}
where ${\Omega}_1(P_J)$ and ${\Omega}_2(P_J)$ are shown below to be positive and strictly monotonically decreasing and increasing functions, respectively, w.r.t. $P_J$, i.e., ${\Omega}_1^{'}(P_J) < 0$ and ${\Omega}_2^{'}(P_J) > 0$ for any $P_J\geq 0$. We will apply $x=\frac{1}{P_J}$ where $x \in (0,\infty)$ as $P_J \in (0, \infty)$. Now, recall $f_e=(\frac{d_{AE_e}}{d_{BE_e}})^\alpha \frac{P_J}{P_T}$ and ${g}=\frac{\beta(1+\rho P_J|g_B|^2)}{P_T\|\mathbf{h}\|^2}$. Then, it follows that
\begin{eqnarray} \small
{\Omega}_1(x)&=& {e^{x d_{BE_e}^\alpha}}\left(1-\frac{k_e}{x+k_e+\rho|g_B|^2} \right) \nonumber\\
{\Omega}_1^{'}(x)&=& {\Omega}_1(x)\left( d_{BE_e}^\alpha+ \frac{k_e}{(x+k_e+\rho|g_B|^2)(x+\rho|g_B|^2)} \right), \nonumber \\
\end{eqnarray}
where $k_e=\left(\frac{d_{AEe}}{d_{BEe}}\right)^\alpha\frac{\|\mathbf{h}\|^2}{\beta}$, and ${\Omega}_1(x)$ and ${\Omega}_1^{'}(x)$ are strictly positive. Also,
\begin{eqnarray} \small
{\Omega}_2(x)&=& \frac{1}{\left(1+ \frac{kx}{x+\rho|g_B|^2}\right)^{M-1}} \nonumber\\
{\Omega}_2^{'}(x)&=& -{\Omega}_2(x)\frac{(M-1)k\rho|g_B|^2}{(x+kx+\rho|g_B|^2)(x+\rho|g_B|^2)}, \nonumber \\
\end{eqnarray}
where $k=\frac{\epsilon P_T ||\mathbf{h}||^2}{(M-1)\beta}$, and ${\Omega}_2(x)$ and ${\Omega}_2^{'}(x)$ are strictly positive and negative respectively.

 Next, we will show that ${\Omega}(x)={\Omega}_1(x){\Omega}_2(x)$ is a unimodal function with minimum at a finite nonzero $x$. Consider the following stationary condition on $x$ ${\Omega}^{'}(x)={\Omega}_1(x){\Omega}_2^{'}(x)+{\Omega}_2(x){\Omega}_1^{'}(x)=0$ or equivalently $\frac{{\Omega}_2^{'}(x)}{{\Omega}_2(x)}=-\frac{{\Omega}_1^{'}(x)}{{\Omega}_1(x)}$ which can be further reduced to
\begin{align} \label{eqn grad Omega} \small
\frac{{\Omega}_2^{'}(x)}{{\Omega}_2(x)}&=-\frac{{\Omega}_1^{'}(x)}{{\Omega}_1(x)} \nonumber\\
d_{BE_e}^\alpha +\frac{M}{x+\rho|g_B|^2}&= \frac{1}{x+k_e+\rho|g_B|^2}+\frac{M-1}{x+\frac{\rho|g_B|^2}{k+1}}.
\end{align}
Using $y=x+\rho|g_B|^2$ in \eqref{eqn grad Omega} and after some algebraic manipulations, we get
\begin{align}\label{cubic} \small
&y^3+ \left(k_e -\frac{k\rho|g_B|^2}{k+1}\right)y^2+ \left (\frac{k_e}{d_{BE_e}^\alpha}-\frac{k\rho|g_B|^2}{k+1}(k_e \right . \notag\\
&\left . +\frac{M-1}{d_{BE_e}^\alpha})\right )y
 -\frac{Mk_e k\rho|g_B|^2}{d_{BE_e}^\alpha (k+1)}=0,
\end{align}
which is a cubic polynomial equation. Based on the characteristics of cubic polynomials, \eqref{cubic} has one, two or three roots and one inflection point \cite{Cubic}. Furthermore, a cubic function is anti-symmetric around its inflection point. To show that \eqref{cubic} has only one positive solution, we just need to show that the inflection point is negative. The inflection point is where the second-order derivative of the cubic function is zero, i.e., $6y-2(k_e -\frac{k\rho|g_B|^2}{k+1})=0$, or equivalently $x=-\frac{2k+3}{3(k+1)}\rho|g_B|^2-\frac{k_e}{3}$, which in this case is indeed negative.

Finally, it is easy to verify that $\Omega(\frac{1}{g};r,\theta)$ is a decreasing function of $P_J$ at $P_J=0$. Therefore, we have shown that $\Omega(\frac{1}{g};r,\theta)$ for any $r$ and $\theta$ has its minimum at a positive finite $P_J$.
\section{Proof of \eqref{eq:TAB:alpha=2}} \label{sec:TAB_alpha=2}
Assume $\alpha=2$ and $\beta=1$. Then, $g=1/z$, and
\begin{align}\label{sigma1} \small
&\int_{0}^{R}\int_{0}^{2\pi}\Omega(z;r,\theta)\mbox{d}\theta r \mbox{d}r\nonumber\\
&=\frac{1}{(1+\frac{z\epsilon}{M-1})^{M-1}}\int_{0}^{R}\int_{0}^{2\pi}\frac{1}{1+zm\frac{r^2}{r^2+d^2-2rdcos\theta}}\mbox{d}\theta r\mbox{d}r\nonumber\\
&=\frac{1}{(1+\frac{z\epsilon}{M-1})^{M-1}}\int_{0}^{R}\int_{0}^{2\pi}\big(1-\nonumber\\
&\qquad \frac{zmr^2}{(1+zm)r^2+d^2-2rdcos\theta}\big)\mbox{d}\theta r\mbox{d}r,
\end{align}
where
\begin{align}\label{sigma2} \small
&\int_{0}^{2\pi}\big(1-\frac{zmr^2}{(1+zm)r^2+d^2-2rdcos\theta}\big)\mbox{d}\theta\nonumber\\
&=2\pi\bigg(1-\frac{1}{\sqrt{1+\frac{(r+d)^2}{r^2zm}}\sqrt{1+\frac{(r-d)^2}{r^2zm}}}\bigg).
\end{align}
Combining  \eqref{sigma1} and \eqref{sigma2} yields (27).
\section{Proof of \eqref{eq:HD_TAB1}} \label{sec:HD_TAB}
Assuming $P_J=0$ and a large $P_T$, it follows that $c=\frac{\|\mathbf{h}\|^2}{\beta}-\frac{1-\frac{1}{\beta}}{(1-\epsilon)P_T}\approx \frac{\|\mathbf{h}\|^2}{\beta}$ and $z=P_T \|\mathbf{h}\|^2$. And from (\ref{TAB_perf}), we have
\vspace{-.05in}
\begin{eqnarray} \small
&&P[S_{AB}>R_S|\Phi,\mathbf{h},g_B]\nonumber\\
&=&\prod_{e\in\Phi}P\big[\frac{X_1\Theta}{d_{AE_e}^\alpha+\frac{\epsilon P_T}{M-1}X_1(1-\Theta)}<{c}|\Phi,\mathbf{h},g_B\big]\nonumber\\
&=&\prod_{e\in\Phi}P\big[\frac{X_4}{d_{AE_e}^\alpha+\frac{\epsilon P_T}{M-1}X_{4,4}}<\frac{\|\mathbf{h}\|^2}{\beta}|\Phi,\mathbf{h},g_B\big],
\end{eqnarray}
where $X_4=\Theta X_1$ is exponentially distributed with mean =1 and $X_{4,4}=(1-\Theta)X_1$ is independent of $X_4$ and has the $\Gamma(M-1,1)$ distribution. Then,
\vspace{-.05in}
\begin{align} \small
&P\big[\frac{X_4}{d_{AE_e}^\alpha+\frac{\epsilon P_T}{M-1}X_{4,4}}<\frac{\|\mathbf{h}\|^2}{\beta}|\Phi,\mathbf{h},g_B\big]\nonumber\\
&=P\big[X_4<\left(\frac{\|\mathbf{h}\|^2}{\beta}d_{AE_e}^\alpha+\frac{\|\mathbf{h}\|^2}{\beta}\frac{\epsilon P_Ty}{M-1}\right)|\Phi,\mathbf{h},g_B\big]\nonumber \\
&=\int_{0}^{\infty}F_{X_4}\left(\frac{\|\mathbf{h}\|^2}{\beta}d_{AE_e}^\alpha+\frac{\|\mathbf{h}\|^2}{\beta}\frac{\epsilon P_Ty}{M-1}\right)f_{X_{4,4}}(y)\mbox{d}y\nonumber\\
&=\int_{0}^{\infty}\left[1-\exp\left\{-\left(\frac{\|\mathbf{h}\|^2}{\beta}d_{AE_e}^\alpha+\frac{\|\mathbf{h}\|^2}{\beta}\frac{\epsilon P_Ty}{M-1}\right)\right\}\right] \nonumber\\
&\times f_{X_{4,4}}(y)\mbox{d}y\nonumber\\
&=1-\int_{0}^{\infty}\exp\left\{-\left(\frac{\|\mathbf{h}\|^2}{\beta}d_{AE_e}^\alpha+\frac{\|\mathbf{h}\|^2}{\beta}\frac{\epsilon P_Ty}{M-1}\right)\right\}\nonumber \\
&\times \frac{y^{M-2}e^{-y}}{\Gamma\left(M-1\right)}\mbox{d}y\nonumber\\
&=1-\frac{e^{-\frac{\|\mathbf{h}\|^2}{\beta}d_{AE_e}^\alpha}}{\Gamma(M-1)}\int_{0}^{\infty}e^{-(1+\frac{\|\mathbf{h}\|^2}{\beta}\frac{\epsilon P_T}{M-1})y}y^{M-2}\mbox{d}y\nonumber\\
&=1-\frac{e^{-\frac{z}{\beta}\frac{d_{AE_e}^\alpha}{P_T}}}{(1+\frac{z}{\beta}\frac{\epsilon}{M-1})^{M-1}}\\
&=1-\Omega(z;r,\theta),
\end{align}
And then
\begin{align}\label{eq:HD_TAB} \small
&\int_{0}^{R}\int_{0}^{2\pi}\Omega(z;r,\theta)d\theta r\mbox{d}r =\frac{2\pi}{(1+\frac{\epsilon z}{\beta(M-1)})^{M-1}}\int_{0}^{R}e^{-\frac{z r^\alpha}{\beta P_T}}r\mbox{d}r\nonumber\\
&=\frac{2\pi{\beta}^{\frac{2}{\alpha}}}{\alpha(\|\mathbf{h}\|^2)^{\frac{2}{\alpha}}(1+\frac{\epsilon P_T}{M-1}\frac{\|\mathbf{h}\|^2}{\beta})^{M-1}}\int_{0}^{\frac{R^{\alpha}\|\mathbf{h}\|^2}{\beta}}e^{-y}y^{\frac{2}{\alpha}-1}y\mbox{d}y\nonumber\\
&=\frac{2\pi{\beta}^{\frac{2}{\alpha}}}{\alpha(\|\mathbf{h}\|^2)^{\frac{2}{\alpha}}(1+\frac{\epsilon P_T}{M-1}\frac{\|\mathbf{h}\|^2}{\beta})^{M-1}}\mathbf{\gamma}\big(\frac{2}{\alpha},\frac{R^{\alpha}
\|\mathbf{h}\|^2}{\beta}\big).
\end{align}
which is \eqref{eq:HD_TAB1}.
\section{Proof of \eqref{eq:ordering}} \label{sec:TAB_ordering}
\begin{align} \small
&P[S_{AB_n}>R_s|d_{AB_n},\Phi_E]\nonumber\\
&=P[\max\limits_{e\in\Phi_E}SNR_{AE_e}<\frac{SNR_{AB_n}}{\beta}+(\frac{1}{\beta}-1)]\nonumber \\
&=P[\max\limits_{e\in\Phi_E}\frac{(1-\epsilon)P_{T}X_4}{ d_{AE_e}^\alpha+\frac{\epsilon P_T}{M-1}X_{4,4}}<\frac{(1-\epsilon)P_TX_{2,n} }{\beta{d_{AB_n}^\alpha} }+(\frac{1}{\beta}-1)]\nonumber \\
&\approx P[\max\limits_{e\in\Phi_E}\frac{X_4}{d_{AE_e}^\alpha+\frac{\epsilon P_T}{M-1}X_{4,4}}<\frac{X_{2,n}}{\beta{d_{AB_n}^\alpha}}]\text{\;\;\;\;(for large $P_T$)}\nonumber
\end{align}
And then
\begin{align} \small
&P[\max\limits_{e\in\Phi_E}\frac{X_4}{d_{AE_e}^\alpha+\frac{\epsilon P_T}{M-1}X_{4,4}}<\frac{X_{2,n}}{\beta{d_{AB_n}^\alpha}}]\nonumber\\
&=\prod\limits_{e\in\Phi_E}P[\frac{X_4}{X_{2,n}}<\frac{d_{AE_e}^\alpha+\frac{\epsilon P_T}{M-1}X_{4,4}}{\beta d_{AB_n}^\alpha}]\nonumber\\
&=\prod\limits_{e\in\Phi_E}\int_{0}^{\infty}f_{X_{4,4}}(x)F_{\frac{X_4}{X_{2,n}}}\left(\frac{d_{AE_e}^\alpha+\frac{\epsilon P_T}{M-1}X_{4,4}}{\beta d_{AB_n}^\alpha}\right)\mbox{d}x\nonumber\\
&=\prod\limits_{e\in\Phi_E}\int_{0}^{\infty}f_{X_{4,4}}(x)\left[1-\left(1+\frac{d_{AE_e}^\alpha+\frac{\epsilon P_T}{M-1}X_{4,4}}{\beta d_{AB_n}^\alpha}\right)^{-M}\right]\mbox{d}x\nonumber\\
&=\prod\limits_{e\in\Phi_E}\left(1-\int_{0}^{\infty}f_{X_{4,4}}(x)\left(1+\frac{d_{AE_e}^\alpha+\frac{\epsilon P_T}{M-1}x}{\beta d_{AB_n}^\alpha}\right)^{-M}\right)\nonumber\\
&=\prod\limits_{e\in\Phi_E}\big(1-\frac{1}{(\frac{\epsilon P_T}{(M-1)\beta d_{AB_n}^\alpha})^M}U(M,2,\frac{(M-1)\beta d_{AB_n}^\alpha}{\epsilon P_T}\nonumber\\
&\qquad +\frac{(M-1)d_{AE_e}^\alpha}{\epsilon P_T})\big),
\end{align}
where $U$ denotes the confluent hypergeometric function of the second kind \cite{Confluent}. After applying Campbell's theorem \cite{campbell} and setting $d_{AE_e}=r$, we have
\begin{align} \small
&P[S_{AB}>R_s|d_{AB_n}]=\exp\big(-2\pi\rho_E {(\frac{(M-1)\beta d_{AB_n}^\alpha}{\epsilon P_T})^M}\nonumber \\
&\times\int_{0}^{\infty}U(M,2,\frac{(M-1)\beta d_{AB_n}^\alpha}{\epsilon P_T}+\frac{(M-1)r^\alpha}{\epsilon P_T})r\mbox{d}r\big).
\end{align}
Further simplification can be done by using proof as shown in subsection \ref{sec:ordering_U}:
\vspace{-.05in}
\begin{align}\label{eq:40b} \small
   & {\left(\frac{(M-1)\beta d_{AB_n}^\alpha}{\epsilon P_T}\right)^M}\int_{0}^{\infty}U(M,2,\frac{(M-1)\beta d_{AB_n}^\alpha}{\epsilon P_T}+\notag\\
   &\qquad \frac{(M-1)r^\alpha}{\epsilon P_T})r\mbox{d}r \notag\\
   & =\frac{1}{\alpha}\frac{({\beta d_{AB_n}^\alpha})^MB(M-\frac{2}{\alpha},\frac{2}{\alpha})}{(\frac{\epsilon P_T}{M-1})^{M-\frac{2}{\alpha}}}U(M-\frac{2}{\alpha},2-\frac{2}{\alpha},\notag\\
   &\qquad \frac{(M-1)\beta d_{AB_n}^\alpha}{\epsilon P_T}).
\end{align}
\subsection{Proof of \eqref{eq:40b}} \label{sec:ordering_U}
Using the change of variables $c=\frac{(M-1)\beta d_{AB_n}^\alpha}{\epsilon P_T}$, we can write
\begin{align} \small
&{(\frac{(M-1)\beta d_{AB_n}^\alpha}{\epsilon P_T})^M}\int_{0}^{\infty}U(M,2,\frac{(M-1)\beta d_{AB_n}^\alpha}{\epsilon P_T}\nonumber\\
&\qquad+\frac{(M-1)r^\alpha}{\epsilon P_T})r\mbox{d}r\nonumber\\
&=\frac{c^M}{\Gamma(M)}\int_{r=0}^{\infty}\int_{t=0}^{\infty}e^{-(c+\frac{(M-1)r^\alpha}{\epsilon P_T})t}t^{M-1}(1+t)^{2-M-1}\mbox{d}tr\mbox{d}r\nonumber\\
&=\frac{c^M}{\Gamma(M)}\int_{t=0}^{\infty}\bigg(e^{-ct}t^{M-1}(1+t)^{1-M}\int_{r=0}^{\infty}e^{-\frac{(M-1)r^\alpha}{\epsilon P_T}t}r\mbox{d}r\bigg)\mbox{d}t.\nonumber
\end{align}
Using another change of variables $x=\frac{(M-1)r^\alpha}{\epsilon P_T}t$, the above becomes
\begin{align} \small
&\frac{(\frac{\epsilon P_T}{M-1})^\frac{2}{\alpha}c^M\Gamma(\frac{2}{\alpha})}{\alpha\Gamma(M)}\int_{0}^{\infty}
e^{-ct}t^{M-\frac{2}{\alpha}-1}\notag\\
&\qquad \times (1+t)^{2-\frac{2}{\alpha}-M+\frac{2}{\alpha}-1}\mbox{d}t\nonumber\\
&=\frac{(\frac{\epsilon P_T}{M-1})^\frac{2}{\alpha}c^M\Gamma(\frac{2}{\alpha})\Gamma(M-\frac{2}{\alpha})}{\alpha\Gamma(M)}\frac{1}{\Gamma(M-\frac{2}{\alpha})}\nonumber\\
&\qquad \times\int_{0}^{\infty}e^{-ct}t^{M-\frac{2}{\alpha}-1}(1+t)^{2-\frac{2}{\alpha}-M+\frac{2}{\alpha}-1}\mbox{d}t\nonumber\\
&=\frac{(\frac{\epsilon P_T}{M-1})^\frac{2}{\alpha}c^MB(M-\frac{2}{\alpha},\frac{2}{\alpha})}{\alpha}\nonumber\\
&\qquad \times U(M-\frac{2}{\alpha},2-\frac{2}{\alpha},\frac{(M-1)\beta d_{AB_n}^\alpha}{\epsilon P_T}).
\end{align}

\section{Proof of \eqref{ep=0_UE}} \label{sec:TAB_ordering0}
When $\epsilon=0$, we have
\begin{eqnarray} \small
&&P[S_{AB_n}>R_s|d_{AB_n},\Phi_E]\nonumber\\
&=&P[\max\limits_{e\in\Phi_E}\frac{P_{T}X_4}{d_{AE_e}^\alpha}<\frac{P_TX_{2,n}}{\beta{d_{AB_n}^\alpha}}+\frac{1}{\beta}-1]\nonumber \\
&\approx&P[\max\limits_{e\in\Phi_E}\frac{X_4}{d_{AE_e}^\alpha}<\frac{X_{2,n}}{\beta{d_{AB_n}^\alpha}}] \text{\;\;\;\;(for large $P_T$)}\nonumber\\
&=&\prod\limits_{e\in\Phi_E}P[\frac{X_4}{X_{2,n}}<\frac{d_{AE_e}^\alpha}{\beta d_{AB_n}^\alpha}],
\end{eqnarray}
where $\frac{X_4}{X_{2,n}}$ follows an F-distribution, i.e., $f_{\frac{X_4}{X_{2,n}}}(x)=M(1+x)^{-(M+1)}$ and $F_{\frac{X_4}{X_{2,n}}}(x)=1-(1+x)^{-M}$. Then,
\begin{align} \small
P[S_{AB_n}>R_s|d_{AB_n},\Phi_E]
&=\prod\limits_{e \in \Phi_E}\big(1-(1+\frac{d_{AE_e}^\alpha}{\beta d_{AB_n}^\alpha})^{-M}\big).
\end{align}
After applying the Campbell's theorem \cite{campbell} and setting $d_{AE_e}=r$ and $d_{AB_n}=x$, we have 
\begin{align}\label{eq:PSABn} \small
P[S_{AB_n}>R_s|d_{AB_n}]
&=\exp\bigg(-\frac{2}{\alpha}\pi\rho_E{\beta}^{\frac{2}{\alpha}}d_{AB_n}^2 \nonumber\\
&\qquad \times B(M-\frac{2}{\alpha},
\frac{2}{\alpha})\bigg).
\end{align}
The computation of averaged SOP requires the PDF of the distance of the $n$th nearest user. The following lemma is known \cite{G.chen_ordering}:
\begin{lemma}\label{d_AB}
	The PDF of $d_{AB_n}$ is
	\begin{equation}
	f_{d_{AB_n}}(x)=\exp(-\rho_U\pi x^2)\frac{2\rho_U^n\pi^nx^{2n-1}}{\Gamma (n)}.
	\end{equation}
\end{lemma}
It now follows from this lemma and \eqref{eq:PSABn} that
\begin{align} \small
P[S_{AB}>R_s]
&=\int_{0}^{\infty} \exp \bigg( -\frac{2}{\alpha}\pi\rho_E {\beta}^{\frac{2}{\alpha}} x^2 B(M-\frac{2}{\alpha},\frac{2}{\alpha})\bigg) \nonumber\\
&\qquad \times \exp(-\rho_U\pi x^2) \frac{2 \rho_U^n \pi^n x^{2n-1}}{\Gamma (n)} \mbox{d}x\nonumber\\
&=\frac{1}{\bigg(1+\frac{\rho_E}{\rho_U}\frac{2}{\alpha}\beta^{\frac{2}{\alpha}}
B(M-\frac{2}{\alpha},\frac{2}{\alpha}) \bigg)^n}. \nonumber
\end{align}
\section{Proof of \eqref{eq:laplace_FD_TAS}} \label{sec:Laplace}

By definition, we have
\begin{eqnarray} \small
\mathcal{L}_{I_e}(s)&=&E_{\Phi}E_{I_e}\big[\exp(-s\sum\limits_{e\in\Phi}\frac{|h_{A_{i^*}E_e}|^2}{\frac{d_{AE_e}^{\alpha}}{P_T}+\frac{d_{AE_e}^{\alpha}}{d_{BE_e}^{\alpha}}mX_2})\big]\nonumber\\
&=&E_{\Phi}\big[\prod\limits_{e\in\Phi}E[\exp(-sH_e)]\big],
\end{eqnarray}
where $H_e=\frac{1}{f_e}\frac{|h_{A_{i^*}E_e}|^2}{\frac{d_{BE_e}^{\alpha}}{P_J}+X_2}=
\frac{1}{f_e}\mathcal{X}_e$. From Lemma \ref{Lemma1} or \eqref{eq:18} with $M=1$, we know $f_{\mathcal{X}_e}(x_e)=e^{-\frac{d_{BE_e}^\alpha}{P_J}x_e}\left(\frac{1}{(1+x_e)^2}+
\frac{\frac{d_{BE_e}^\alpha}{P_J}}{1+x_e}\right)$. Then,
\begin{align} \small
&E[e^{-sH_e}]=\int_{0}^{\infty}e^{-sh_e}f_{H_e}(h_e)dh_e =\int_{0}^{\infty}e^{-s\frac{x_e}{f_e}}f_{\mathcal{X}_e}(x_e)\mbox{d}x_e\nonumber\\
&=\int_{0}^{\infty}e^{-\frac{(s+\frac{d_{AE_e}^{\alpha}}{P_T})}{f_e}x_e}\left(\frac{1}{(1+x_e)^2}+
\frac{d_{BE_e}^{\alpha}}{P_J}\frac{1}{(1+x_e)}\right)\mbox{d}x_e\nonumber\\
&=\int_{0}^{\infty}e^{-K(s)x_e}\left(\frac{1}{(1+x_e)^2}+(K(s)-\frac{s}{f_e})\frac{1}{(1+x_e)}\right)
\mbox{d}x_e\nonumber\\
&=\int_{0}^{\infty}\frac{e^{-K(s)x_e}}{(1+x_e)^2}\mbox{d}x_e+(K(s)-\frac{s}{f_e})\int_{0}^{\infty}
\frac{e^{-K(s)x_e}}{(1+x_e)}\mbox{d}x_e\nonumber\\
&=1-e^{K(s)}K(s)\mathbf{E_1}(K(s))+(K(s)-\frac{s}{f_e})e^{K(s)}\mathbf{E_1}(K(s))\nonumber\\
&=1-\frac{s}{f_e}\mathbf{E_1}(K(s))e^{K(s)},
\end{align}
where $\mathbf{E_1}(a)=\int_0^\infty\frac{e^{-ax}}{1+x}\mbox{d}x$ and $K(s)=\frac{s+\frac{d_{AE_e}^{\alpha}}{P_T}}{f_e}$. So we get:
\begin{align} \small
\mathcal{L}_{I_e}(s)&=E_{\Phi}\big[\prod\limits_{e\in\Phi}E[\exp(-sH_e)]\big]\nonumber\\
&=E_{\Phi}\big[\prod\limits_{e\in\Phi}1-\frac{s}{f_e}\mathbf{E_1}(K(s))e^{K(s)}\big]\nonumber\\
&=\exp\bigg[-\rho_E\int_{0}^{R}\int_{0}^{2\pi}\frac{s}{f_e}\mathbf{E_1}(K(s))e^{K(s)}\mbox{d}\theta r\mbox{d}r\bigg].
\end{align}

\begin{IEEEbiography}
[{\includegraphics[width=1in,height=1.25in,clip,keepaspectratio]{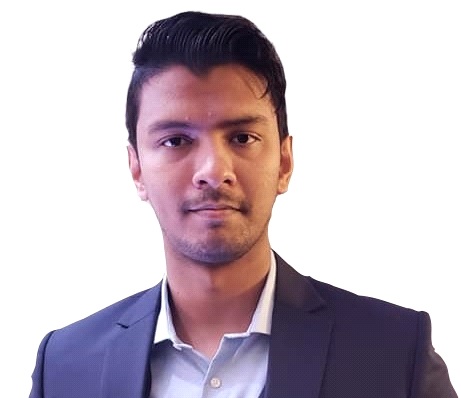}}]{Ishmam Zabir}
(Student Member, IEEE) received the B.Eng. Degree in electrical and electronics engineering from Bangladesh University of Engineering and Technology, Bangladesh, in 2015. He is currently working towards the PhD degree in electrical and computer engineering at university of California Riverside, Riverside, CA, USA. His research interests include physical layer security in wireless network, full-duplex radio, resource allocation in distributed network and stochastic geometry.
\end{IEEEbiography}
\vspace{-.1in}
\begin{IEEEbiography}
[{\includegraphics[width=1in,height=1.25in,clip,keepaspectratio]{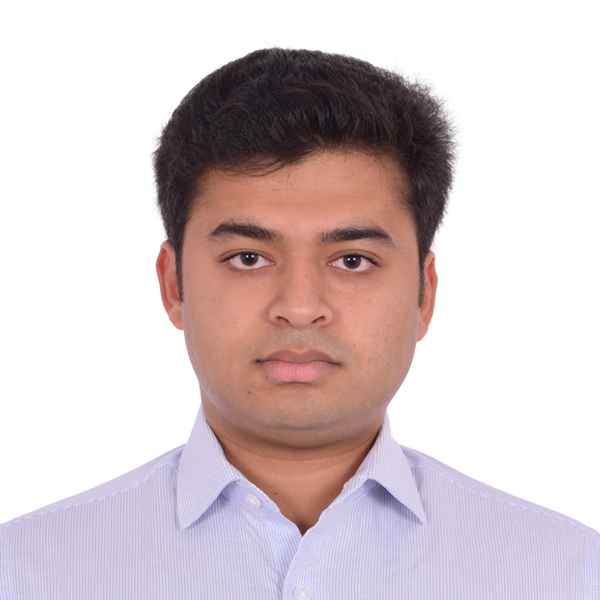}}]{Ahmed Maksud}
 (Student Member, IEEE) received the B.Sc. degree in Electrical and Electronics Engineering from Bangladesh University of Engineering and Technology, Dhaka, Bangladesh in 2017. He is currently working towards the Ph.D. degree in Electrical and Computer Engineering at the University of California at Riverside, CA, USA. His research interests include Wireless Communication, Wireless Networks Security, and Full-duplex Radio.
\end{IEEEbiography}
\vspace{-.1in}
\begin{IEEEbiography}
[{\includegraphics[width=1in,height=1.25in,clip,keepaspectratio]{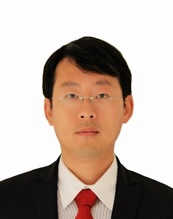}}]{Gaojie Chen} (S'09 -- M'12 -- SM'18) received the B.Eng. and B.Ec. Degrees in electrical information engineering and international economics and trade from Northwest University, China, in 2006, and the M.Sc. (Hons.) and Ph.D. degrees in electrical and electronic engineering from Loughborough University, Loughborough, U.K., in 2008 and 2012, respectively. He is currently a Lecturer with the School of Engineering, University of Leicester, U.K. He serves as an Associate Editor for the {\scshape IEEE Communications Letters}, {\scshape IEEE Wireless Communications Letters}, {\scshape IEEE Journal on Selected Areas in Communications - Machine Learning in Communications and Networks} and {\it Electronics Letters} (IET).
\end{IEEEbiography}
\vspace{-.1in}
\begin{IEEEbiography}
[{\includegraphics[width=1in,height=1.25in,clip,keepaspectratio]{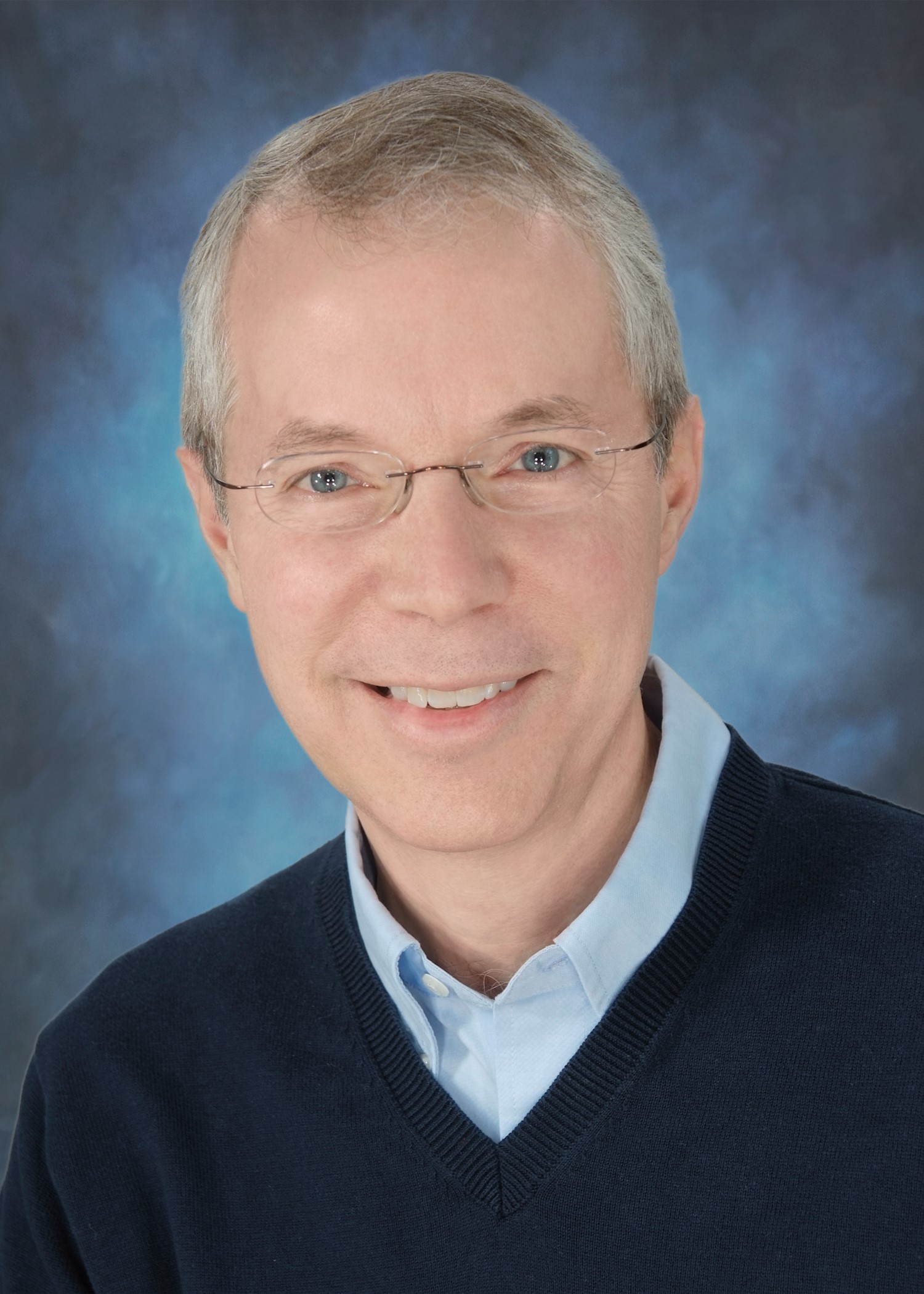}}]{Brian M. Sadler} (S'81--M'81--SM'02--F'07) received the B.S. and M.S. degrees from the University of Maryland, College Park, and the PhD degree from the University of Virginia, Charlottesville, all in electrical engineering.  He is the US Army Senior Scientist for Intelligent Systems, and a Fellow of the Army Research Laboratory (ARL) in Adelphi, MD.  He is an IEEE Communications Society Distinguished Lecturer for 2020-2021, was an IEEE Signal Processing Society Distinguished Lecturer for 2017-2018, and general co-chair of IEEE GlobalSIP’16.  He has been an Associate Editor of the IEEE Transactions on Signal Processing, IEEE Signal Processing Letters, and EURASIP Signal Processing, and a Guest Editor for several journals including the IEEE JSTSP, the IEEE JSAC, IEEE T-RO, the IEEE SP Magazine, Autonomous Robots, and the International Journal of Robotics Research. He received Best Paper Awards from the IEEE Signal Processing Society in 2006 and 2010, several ARL and Army R/D awards, and a 2008 Outstanding Invention of the Year Award from the University of Maryland.  He has over 400 publications in these areas with 17,000 citations and h-index of 56. His research interests include information science, and networked collaborative autonomous intelligent systems.
\end{IEEEbiography}
\vspace{-.1in}
\begin{IEEEbiography}
[{\includegraphics[width=1in,height=1.25in,clip,keepaspectratio]{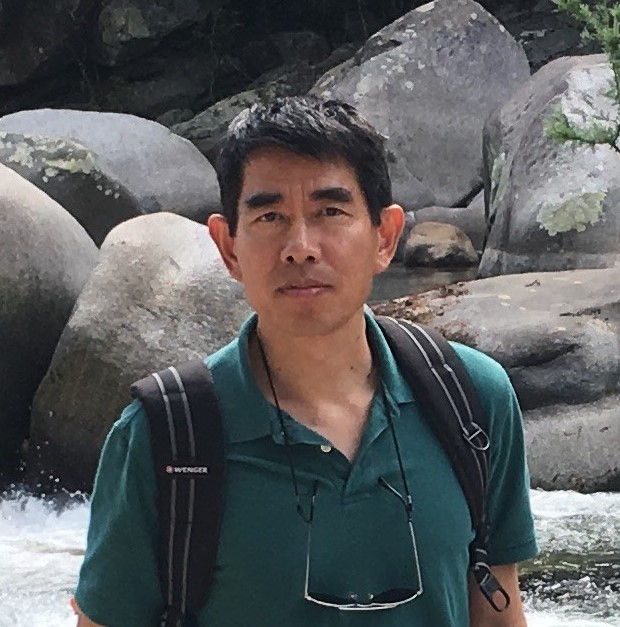}}]{Yingbo Hua} (S'86--M'88--SM'92--F'02) received a B.S. degree in 1982 from Southeast University, Nanjing, China, and a Ph.D. degree in 1988 from Syracuse University, NY, USA. He joined the faculty of the University of Melbourne, Australia, in 1990, where he was promoted to Reader and Associate Professor in 1996. He moved to the University of California, Riverside, CA, as Full Professor in 2001, where he advanced to Senior Full Professor in 2009 and Professor IX in 2019. Dr. Hua has published over 340 articles in signal processing, wireless communications and sensor networks, with citations over 13520 and h-Index 55. Since 1994, he has served in a number of editorial and leadership roles for IEEE Transactions on Signal Processing, IEEE Signal Processing Letters, Signal Processing, IEEE Signal Processing Magazine, IEEE Journal of Selected Areas in Communications, IEEE Wireless Communications Letters, and IEEE Transactions on Signal and Information Processing over Networks. He was a General Co-Chair for IEEE ChinaSIP’2015, and Lead Chair for IEEE GlobalSIP’2018 Symposium on Signal Processing for Wireless Network Security. He is currently Chair of the Steering Committee for IEEE Wireless Communications Letters. Dr. Hua is a Fellow of AAAS since 2011.
\end{IEEEbiography}


\begin{thebibliography}{99}

\bibitem{wyner}
A. D. Wyner, ``The wire-tap channel'', Bell Syst. Tech. J., vol. 54, pp. 1355-1387, Jan. 1975.

\bibitem{yang}
N. Yang, L. Wang, G. Geraci, M. Elkashlan, J. Yuan, and M. D. Renzo, ``Safeguarding 5G wireless communication networks using physical layer security,'' IEEE Commun. Mag., vol. 53, no. 4, pp. 20-27, Apr. 2015.

\bibitem{tekin}
E. Tekin and A. Yener, ``The general gaussian multiple access and twoway wire-tap channels: achievable rates and cooperative jamming,'' IEEE Trans. Inform. Theory, vol. 54, pp. 2735-2751, June 2008.

\bibitem{zhang}
H. Zhang, H. Xing, J. Cheng, A. Nallanathan, and V. Leung, ``Secure resource allocation for OFDMA two-way relay wireless sensor networks without and with cooperative jamming,'' IEEE Trans. Ind. Informat., vol. 12, no. 5, pp. 1714-1725, Oct. 2016.


\bibitem{Zhang2010}
J. Zhang and M. C. Gursoy, ``Relay beamforming strategies for physicallayer security,'' in Proc. 44th Annu. Conf. Inf. Sci. Syst., Princeton, NJ, USA, Mar. 2010, pp. 1-6.

\bibitem{weingarten}
H. Weingarten, T. Liu, S. Shamai, Y. Steinberg, and P. Viswanath, ``The capacity region of the degraded multiple-input multiple-output compound broadcast channel,'' IEEE Trans. Inf. Theory, vol. 55, pp. 5011-5023, Nov. 2009.

\bibitem{Chen2017}
L. Chen, Q. Zhu, W. Meng and Y. Hua, ``Fast Power Allocation for Secure Communication with Full-Duplex Radio'', IEEE Trans. Signal Processing, Vol. 65, No. 14, pp. 3846-3861, July 2017.

\bibitem{Chen2016}
G. Chen, Y. Gong, P. Xiao, and J. A. Chambers, ``Dual antenna selection in secure cognitive radio networks,'' IEEE Trans. Veh. Technol., vol. 65, no. 10, pp. 7993-8002, Oct. 2016.

\bibitem{khisti}
A. Khisti, G. Wornell, A. Wiesel, and Y. Eldar, ``On the Gaussian MIMO wiretap channel,'' in Proc. IEEE Int. Symp. Information Theory, Nice, France, June 2007.

\bibitem{goel}
S. Goel and R. Negi, ``Guaranteeing secrecy using artificial noise,'' IEEE Trans. Wireless Commun., vol. 7, no. 6, pp. 2180-2189, June 2008.

\bibitem{Huang2016}
Y. Huang, J. Wang, C. Zhong, T. Q. Duong, and G. K. Karagiannidis, ``Secure transmission in cooperative relaying networks with multiple antennas,'' IEEE Trans. Wireless Commun., vol. 15, no. 10, pp. 6843-6856, Oct. 2016.

\bibitem{Yang2017}
M. Yang, B. Zhang, Y. Huang, N. Yang, D. B. da Costa, and D. Guo, ``Secrecy enhancement of multiuser MISO networks using OSTBC and artificial noise,” IEEE Trans. Veh. Technol., vol. 66, no. 12, pp. 11394-11398, Dec. 2017.

\bibitem{haenggi2008}
M. Haenggi, ``The secrecy graph and some of its properties,'' in Proc. IEEE Int. Symp. Inf. Theory, Toronto, Canada, pp. 539-543, July 2008.

\bibitem{Wang2013}
H. Wang, X. Zhou, and M. C. Reed, ``Physical layer security in cellular networks: A stochastic geometry approach,'' IEEE Trans. Wireless Commun., vol. 12, no. 6, pp. 2776-2787, Jun. 2013.

\bibitem{Zhou2011}
X. Zhou, R. K. Ganti, and J. G. Andrews, ``Secure wireless network connectivity with multi-antenna transmission,'' IEEE Trans. Wireless Commun., vol. 10, no. 2, pp. 425-430, Dec. 2011.

\bibitem{Zhang1}
X. Zhang, X. Zhou and M. R. McKay, ``On the design of artificial-noise aided secure multi-antenna transmission in slow fading channels,'' IEEE Trans. Veh. Technol., vol. 62, no. 5, pp. 2170-2181, Jun. 2013.

\bibitem{zheng2015}
T. X. Zheng, H. M. Wang, J. Yuan, D. Towsley, and M. H. Lee, ``Multi-Antenna transmission with artificial noise against randomly distributed eavesdroppers,'' IEEE Trans. Commun., vol. 63, pp. 4347-4362, Nov. 2015.

\bibitem{L Zhang}
L. Zhang, H. Zhang, D. Wu, and D. Yuan, ``Improving physical layer security for MISO systems via using artificial noise,'' in 2015 IEEE Global Communications Conference (GLOBECOM), 2015, pp. 1-6.

\bibitem{zheng3}
T. X. Zheng, H. M. Wang, ``Optimal Power Allocation for Artificial Noise Under Imperfect CSI Against Spatially Random Eavesdroppers,'' IEEE Trans. Vehicular Technology, vol. 65, pp. 8812-8817, Oct. 2016.

\bibitem{Zhang2013}
X. Zhang, X. Zhou, and M. R. McKay, ``Enhancing secrecy with multiantenna transmission in wireless ad hoc networks,'' IEEE Trans. Inf. Forensics and Security, vol. 8, no. 11, pp. 1802-1814, Nov. 2013.

\bibitem{Ghogho2011}
M. Ghogho and A. Swami, ``Physical-layer secrecy of MIMO communications in the presence of a poisson random field of eavesdroppers,'' in Proc. IEEE ICC, Kyoto, Japan, Jun. 2011, pp. 1-5.

\bibitem{Zhou20112}
 X. Zhou, R. Ganti, J. Andrews, and A. Hj\textit{$	\o$}rungnes, ``On the throughput cost of physical layer security in decentralized wireless networks,'' IEEE Trans. Wireless Commun., vol. 10, no. 8, pp. 2764-2775, Aug. 2011.

\bibitem{geraci}
G. Geraci, S. Singh, J. G. Andrews, J. Yuan, and I. B. Collings, ``Secrecy rates in broadcast channels with confidential messages and external eavesdroppers,'' IEEE Trans. Wireless Commun., vol. 13, pp. 2931-2943, May 2014.

\bibitem{zheng1}
T. X. Zheng, H. M. Wang, and Q. Yin, ``On transmission secrecy outage of a multi-antenna system with randomly located eavesdroppers,'' IEEE Commun. Lett., vol. 18, pp. 1299-1302, Aug. 2014.

\bibitem{riihonen}
T. Riihonen, S. Werner, R. Wichman, ``Mitigation of loopback self-interference in full-duplex MIMO relays'', IEEE Trans. Signal Process., Vol. 59, Dec 2011, 5983-5993.

\bibitem{hua2015}
Y. Hua, Y. Ma, A. Gholian, Y. Li, A. Cirik, P. Liang, ``Radio Self-Interference Cancellation by Transmit Beamforming, All-Analog Cancellation and Blind Digital Tuning,'' Signal Processing, Vol. 108, pp. 322-340, 2015.

\bibitem{Zheng2017}
T. X. Zheng, H. M. Wang, Q. Yang, and M. H. Lee, ``Safeguarding decentralized wireless networks using full-duplex jamming receivers,'' IEEE Trans. Wireless Commun., vol. 16, no. 1, pp. 278-292, Jan. 2017.

\bibitem{Zhang2017}
T. Zhang, Y. Cai, Y. Huang, T. Q. Duong, and W. Yang, ``Secure full-duplex spectrum-sharing wiretap networks with different antenna reception schemes,'' IEEE Trans. Commun., vol. 65, no. 1, pp. 335-346, Jan. 2017.

\bibitem{Hua2017}
Y. Hua, Q. Zhu, and R. Sohrabi, ``Fundamental Properties of Full-Duplex Radio for Secure Wireless Communications,'' http://arxiv.org/abs/1711.10001, 2017.

\bibitem{hua2019}
Y. Hua, ``Advanced Properties of Full-Duplex Radio for Securing Wireless Network,'' IEEE Trans. Signal Processing, Vol. 67, pp. 120-135, Jan. 2019.

\bibitem{Sohrabi_2019}
R. Sohrabi, Q. Zhu,  Y. Hua, ``Secrecy analyses of a full-duplex MIMOME network,'' IEEE Transactions on Signal Processing, Vol. 67, No. 23, pp. 5968-5982, Dec. 2019.


\bibitem{Xie}
M. Xie and T.-M. Lok, ``Antenna Selection in RF-Chain-Limited MIMO Interference Networks Under Interference Alignment,'' IEEE Trans. on Vehicular Technology, vol. 66, no. 5, pp. 3856-3870,May 2017.

\bibitem{gaojie2017}
G. Chen, J. P. Coon and M. D. Renzo, ``Secrecy Outage Analysis for Downlink Transmissions in the Presence of Randomly Located Eavesdroppers,''IEEE Trans. Information Forensics and Security, Vol. 12, PP. 1195 - 1206, May 2017.



\bibitem{Ishmam2019}

I. Zabir, A. Maksud, B. Sadler and Y. Hua, ``Secure Downlink Transmission to Full-Duplex User against Randomly Located Eavesdroppers,'' IEEE Global Communications Conference (GLOBECOM), Waikoloa, HI, USA, Dec. 2019.

\bibitem{G.chen_ordering}
G. Chen and J. P. Coon, ``Secrecy Outage Analysis in Random Wireless Networks With Antenna Selection and User Ordering,'' IEEE Wireless Commun. Letters, Vol. 6, No. , pp.334-337, June 2017.

\bibitem{Hua2012}
Y. Hua, P. Liang, Y. Ma, A. Cirik and Q. Gao, ``A method for broadband full-duplex MIMO radio,'' IEEE Signal Processing Letters, Vol. 19, No. 12, pp. 793-796, Dec 2012.


%


\bibitem{Fdist}
P. B. Patnaik, ``The non-central $\chi ^2$- and F-distribution and their applications,'' Biometrika, vol. 36, no. 1/2, pp. 202-232, Jun. 1949.

\bibitem{campbell}
R. L. Streit, The Poisson Point Process,  Springer, Boston, MA, 2010.

\bibitem{Ju2018}
Y. Ju, H. M. Wang, T. X. Zheng, Q. Yin, M. H. Lee, ``Safeguarding Millimeter Wave Communications
Against Randomly Located Eavesdroppers'', IEEE Trans. Wireless Commun., Vol. 17, Issue: 4, pp. 2675 - 2689, April 2018.

\bibitem{Wang2016}
C. Wang, H. M. Wang, ``Physical Layer Security in Millimeter Wave Cellular Networks,'' IEEE Trans. Wireless Commun., Vol. 15, Issue: 8, pp. 5569 - 5585, Aug. 2016.

\bibitem{TBai2015}
T. Bai, R. W. Heath, ``Coverage and Rate Analysis for Millimeter-Wave Cellular Networks,''  IEEE Trans. Wireless Commun.,  Vol. 14 Issue: 2, pp. 1100 - 1114, 2, Feb. 2015.







\bibitem{Confluent}
 L. C. Andrews, ``Special Functions of Mathematics for Engineers,'' SPIE Press, chapter 10, 1998.

\bibitem{Cubic}
L. Bostock,  S. Chandler, and F. S. Chandler, Pure Mathematics 2, Nelson Thornes, 1979.


\bibitem{Sanayei2004}
S. Sanayei and A. Nosratinia, ``Antenna Selection in MIMO Systems,'' IEEE Communications Magazine, October 2004.

\bibitem{Garcia2010}
L. Dritsoula, Z. Wang, H. R. Sadjadpour, J.J. Garcia-Luna-Aceves, ``Antenna selection for opportunistic interference management in MIMO broadcast channels,'' IEEE Workshop on Signal Processing Advances in Wireless Communications (SPAWC), 2010.

\bibitem{Gao2016}
Y. Gao, and T. Kaiser, ``Antenna Selection in Massive MIMO Systems:
Full-array Selection or Subarray Selection?'' IEEE Workshop on Sensor Array and Multichannel Signal Processing Workshop, 2016.

\bibitem{Alves2012}
H. Alves,  R. D. Souza, M. Debbah, and M. Bennis, ``Performance of Transmit Antenna Selection
Physical Layer Security Schemes,'' IEEE Signal Processing Letters, June 2012.



\end{thebibliography}
\end{document}